\documentclass[11pt]{article}

\usepackage{amstext,amsgen,latexsym}
\usepackage{amstext,amssymb,amsfonts,latexsym}
\usepackage{theorem}
\usepackage{pifont}

 \newcommand{\bs}{\bigskip}
 \newcommand{\ms}{\medskip}
 \newcommand{\n}{\noindent}
 
 \newcommand{\hs}[1]{\hspace*{ #1 mm}}
 \newcommand{\vs}[1]{\vspace*{ #1 mm}}

 \newcommand{\real}{\mathbb{R}}

 \newcommand{\nat}{\mathbb{N}}
 \newcommand{\dyadic}{\mathbb{D}}
 \newcommand{\integers}{\mathbb{Z}}
 \newcommand{\rational}{\mathbb{Q}}
 
 \newcommand{\complex}{\mathbb{C}}
 \newcommand{\appcomplex}{\tilde{\mathbb{C}}}

 \newcommand{\vsigma}{\mbox{\boldmath $\sigma$}}
 \newcommand{\vtau}{\mbox{\boldmath $\tau$}}
 
 \newcommand{\vepsilon}{\mbox{\boldmath $\epsilon$}}
 \newcommand{\vd}{\mbox{\boldmath $d$}}

 \newcommand{\svepsilon}{\mbox{\boldmath ${}_{\epsilon}$}}
 \newcommand{\svd}{\mbox{\boldmath ${}_{d}$}}

 \newcommand{\co}{\mathrm{co}\mbox{-}}

 \newcommand{\ie}{\textrm{i.e.},\hspace*{2mm}}
 \newcommand{\eg}{\textrm{e.g.},\hspace*{2mm}}
 \newcommand{\etal}{\textrm{et al.}\hspace*{2mm}}

 \newcommand{\CC}{{\cal C}}
 \newcommand{\FF}{{\cal F}}
 
 \newcommand{\HH}{{\cal H}}

 \newcommand{\GG}{{\cal G}}

 \newcommand{\p}{\mathrm{{\bf P}}}
 \newcommand{\np}{\mathrm{{\bf NP}}}

 \newcommand{\bpp}{\mathrm{{\bf BPP}}}

 \newcommand{\pp}{\mathrm{{\bf PP}}}

 \newcommand{\pspace}{\mathrm{{\bf PSPACE}}}

 \newcommand{\dexp}{\mathrm{{\bf EXP}}}

 \newcommand{\app}{\mathrm{{\bf APP}}}
 
 \newcommand{\wpp}{\mathrm{{\bf WPP}}}

 \newcommand{\eqp}{\mathrm{{\bf EQP}}}
 \newcommand{\nqp}{\mathrm{{\bf NQP}}}
 
 \newcommand{\bqp}{\mathrm{{\bf BQP}}}
 \newcommand{\pqp}{\mathrm{{\bf PQP}}}

 \newcommand{\wqp}{\mathrm{{\bf WQP}}}

 \newcommand{\qma}{\mathrm{{\bf QMA}}}

 \newcommand{\qop}{\mathrm{{\bf QOP}}}
 \newcommand{\aqma}{\mathrm{{\bf AQMA}}}
 
 \newcommand{\eqma}{\mathrm{{\bf EQMA}}}

 \newcommand{\fp}{\mathrm{{\bf FP}}}
 
 \newcommand{\fpspace}{\mathrm{\bf FPSPACE}}
 
 \newcommand{\sharpp}{\#\mathrm{{\bf P}}}
 \newcommand{\optp}{\mathrm{{\bf OptP}}}
 \newcommand{\optsharpp}{\mathrm{\bf Opt}\#\mathrm{\bf P}}
\newcommand{\optsharpqp}{\mathrm{\bf Opt}\#\mathrm{\bf QP}}
 
 \newcommand{\gapp}{\mathrm{{\bf GapP}}}

 \newcommand{\feqp}{\mathrm{{\bf FEQP}}}
 
 \newcommand{\sharpqp}{\#\mathrm{{\bf QP}}}
 
 \newcommand{\gapqp}{\mathrm{{\bf GapQP}}}

 \newcommand{\qoptsharpqp}{\mathrm{{\bf Qopt}}\#\mathrm{{\bf QP}}}

 \newcommand{\phelpcomp}[1]%
       {{\mathrm{\bf P}^{#1}_{help}\mbox{-}\mathrm{comp}}}
 \newcommand{\relpcomp}[1]%
       {{\mathrm{\bf P}^{#1}\mbox{-}\mathrm{comp}}}

 \newcommand{\pttsamp}[1]%
       {{\mathrm{\bf P}^{#1}_{\mathrm{tt}}\mbox{-}\mathrm{samp}}}
 \newcommand{\relpsamp}[1]%
       {{\mathrm{\bf P}^{#1}\mbox{-}\mathrm{samp}}}

 \newcommand{\relqpsamp}[1]%
       {{\mathrm{\bf QP}^{#1}\mbox{-}\mathrm{samp}}}

 \newcommand{\relppcomp}[1]%
       {\mathrm{\bf P}_{\mathrm{\bf
P}^{#1}\mbox{-}\mathrm{comp}}^{#1}}
 \newcommand{\relnppcomp}[1]%
       {\mathrm{\bf NP}_{\mathrm{\bf
P}^{#1}\mbox{-}\mathrm{comp}}^{#1}}

 \newcommand{\appin}{\stackrel{\in}{\sim}{}^{p}}
 \newcommand{\stappin}{\stackrel{\in}{\sim}{}^{e}}
 \newcommand{\appsubseteq}{\Subset^{p}}
 \newcommand{\stappsubseteq}{\Subset^{e}}
 \newcommand{\appreduces}{\lessapprox^{p}}

 \newcommand{\low}{\mathrm{{\bf low\mbox{-}}}}

 \newcommand{\IFF}{\Longleftrightarrow}

 \def\bbox{\vrule height6pt width6pt depth1pt}

\theoremstyle{plain}
\theoremheaderfont{\bfseries}
 \newtheorem{theorem}{Theorem}[section]
 \newtheorem{lemma}[theorem]{Lemma}
 \newtheorem{proposition}[theorem]{Proposition}
 \newtheorem{corollary}[theorem]{Corollary}
 \newtheorem{fact}[theorem]{Fact.}

 {\theorembodyfont{\rmfamily}
\newtheorem{definition}[theorem]{Definition}}
 {\theorembodyfont{\rmfamily} }
 {\theorembodyfont{\rmfamily} }

 \newenvironment{proof}{\par \noindent
            {\bf Proof. \hs{2}}}{\hfill$\Box$ \vspace*{3mm}}

 \newenvironment{proofof}[1]{\vspace*{5mm} \par \noindent
         {\bf Proof of #1.\hs{2}}}{\hfill$\Box$ \vspace*{3mm}}

 \newcommand{\floors}[1]{\lfloor #1 \rfloor}
 \newcommand{\pair}[1]{\langle #1 \rangle}

 \newcommand{\qubit}[1]{| #1 \rangle}
 \newcommand{\bra}[1]{\langle #1 |}
 \newcommand{\ket}[1]{| #1 \rangle}
 \newcommand{\measure}[2]{\langle #1 | #2 \rangle}
 
\newif\ifnotesw\noteswtrue
   {\ifnotesw\marginpar[\hfill\(\top\)]{\(\top\)}\fi}%
      {\ifnotesw\marginpar[\hfill\(\bot\)]{\(\bot\)}\fi}
      
\newcommand{\mnote}[1]%
   {\ifnotesw\marginpar%
	  [{\scriptsize\begin{minipage}[t]{\marginparwidth}
	  \raggedleft#1%
		  \end{minipage}}]%
	  {\scriptsize\begin{minipage}[t]{\marginparwidth}
	  \raggedright#1%
		  \end{minipage}}%
    \fi}

\newcommand{\ignore}[1]{}

\begin{document}
\begin{center}
{\large {\bf Quantum Optimization Problems}} \ms\\
{\sc Tomoyuki Yamakami}\footnote{This work is in part 
supported by Canada's NSERC grant.} \ms\\
{\it School of Information Technology and Engineering} \\
{\it University of Ottawa, Ottawa, Ontario, Canada  K1N 6N5} \bs\\
{April 2, 2002}
\end{center}

\pagestyle{plain}


\paragraph{Abstract.}
Krentel [J. Comput. System. Sci., 36, pp.490--509] presented a
framework for an NP optimization problem that searches an optimal
value among exponentially-many outcomes of polynomial-time
computations.  This paper expands his framework to a quantum
optimization problem using polynomial-time quantum computations and
introduces the notion of an ``universal'' quantum optimization problem
similar to a classical ``complete'' optimization problem. We exhibit a
canonical quantum optimization problem that is universal for the class
of polynomial-time quantum optimization problems. We show in a certain
relativized world that all quantum optimization problems cannot be
approximated closely by quantum polynomial-time computations. We also
study the complexity of quantum optimization problems in connection to
well-known complexity classes.
\ms


\section{Introduction}

Quantum computation theory was initiated in the early 1980s and has
shown significant phenomena beyond the classical framework.  During
the 1990s, many classical concepts in complexity theory were examined
and interpreted in quantum context, including the notions of
``bounded-error polynomial-time computation'' \cite{BV97},
``interactive proof system'' \cite{Wat99}, ``parallel query
computation'' \cite{Yam99b,BD99}, ``Kolmogorov complexity''
\cite{Vit01,BDL00}, and ``Merlin-Arthur game''
\cite{Wat00,KMY01}. Along this line of research, this paper studies a
quantum interpretation of a classical optimization problem.

An {\em optimization problem} is in general a certain type of search
problem which is to find, among candidates to which some values are
assigned, the maximal (or minimal) value or to find a solution with
such the value.  A typical example of such an optimization problem is
the {\sc traveling salesperson problem} that asks, upon given a map of
cities and their traveling distances, for the length of a shortest
tour to all the cities in the map.  In the 1980s, Krentel
\cite{Kren88} laid out a framework for studying the complexity of such
optimization problems. He defined $\optp$ to be the collection of
functions outputting the maximal (or minimal) value of indexed functions
computable in polynomial-time. To locate the hardest optimization
problems in $\optp$, Krentel introduced the notion of ``completeness''
under his metric reduction. The {\sc Traveling Salesperson Problem}
turns out to be a complete optimization problem for $\optp$
\cite{Kren88}.
 
This paper studies a quantum interpretation of Krentel's optimization
problem. Krentel's framework is easily generalized to the problem of
optimizing the acceptance probability\footnote{The collection of all
functions that output the acceptance probability of certain
polynomial-time well-formed Turing machines is denoted $\sharpqp$ in
\cite{Yam99b}.} of indexed polynomial-time quantum computations. In
consistence with Krentel's notation, the notation $\optsharpqp$ is
used for the class of these problems. However, this is a mixture of
classical indexing and quantum computations. Instead, we introduce
quantum functions indexed with a quantum state of polynomially-many
qubits (called a {\em quantum index}). A {\em quantum optimization
problem}, discussed in this paper, is to ask for a maximal acceptance
probability of quantum computations labeled with quantum indices. We
use the notation $\qoptsharpqp$ to denote the collection of such
quantum optimization problems.

These quantum optimization problems are also characterized by maximal
eigenvalues of certain types of positive semidefinite, contractive,
Hermitian matrices. In Section \ref{sec:properties}, we use this
characterization to show fundamental properties of quantum
optimization problems in $\qoptsharpqp$.

The existence of {\em complete} optimization problems has largely
contributed to the success of the theory of NP optimization
problems. These complete optimization problems are considered, among
all NP optimization problems, as the ``hardest'' problems to solve. In
Section \ref{sec:universal}, we develop a similar ``hardest'' notion
for quantum optimization problems. As shown by Bernstein and Vazirani
\cite{BV97}, there exists a universal quantum Turing machine that
simulates any well-formed quantum Turing machine with amplitudes
approximable in polynomial time at the cost of polynomial slowdown.
Different from a classical universal Turing machine, since all
amplitudes are only approximated, the universal quantum Turing machine
can simulate other machines only approximately to within a given
closeness parameter. This manner of simulation gives rise to the
notion of approximate reduction. We say that a function $f$ is {\em
approximately reducible to} another function $g$ if $f$ has a
Krentel's metric-type reduction from $f$ to a certain function that
approximates $g$ to within $1/m$, where $m$ is an accuracy parameter
given as an auxiliary input to the reduction. In this fashion, we can
define the notion of {\em universal optimization problems} for the
class $\qoptsharpqp$. This notion requires two functions to be apart
only in a distance that is a reciprocal of a polynomial. In Section
\ref{sec:universal}, we exhibit an
example of a canonical universal optimization problem, which can be
viewed as a generalization of the {\sc Bounded Halting Problem} for
$\np$. This notion of universality naturally induces promise-complete
problems for many well-known quantum complexity classes, such as
$\bqp$ and $\qma$, which are believed to lack complete problems.

The class $\qoptsharpqp$ includes its underlying class $\sharpqp$ as
well as $\optsharpqp$. However, the complexity of $\qoptsharpqp$ is
not well-understood even in comparison with $\sharpqp$. This function
class $\qoptsharpqp$ naturally introduces the class of decision
problems $\qop$ (``quantum optimization polynomial time'') in such a
way that $\pp$ is characterized\footnote{A set $S$ is in $\pp$ iff
there exit two functions $f,g\in\sharpp$ such that, for every $x$,
$x\in A \IFF f(x)>g(x)$.} by two $\sharpp$-functions. This class
$\qop$ lies between $\pp$ and $\pspace$. In Section
\ref{sec:function-class}, we show that, under the assumption
$\eqp=\qop$, every quantum optimization problem in $\qoptsharpqp$ can
be closely approximated by functions in $\sharpqp$. We may not remove
the assumption because there exists a counterexample in a relativized
world. Moreover, if every quantum optimization problem in
$\qoptsharpqp$ can be closely approximated by functions in $\sharpqp$,
then $\qop$ is included in $\p^{\sharpp[1]}$.

The quantum optimization problems are also used to characterize other
major complexity classes. One of the useful tools is the notion of
``definability'', which is adapted from an earlier work of Fenner
\etal \cite{FFK94}. A complexity class $\CC$ is called {\em
$\qoptsharpqp$-definable} if we have a pair of disjoint sets
$A,R\subseteq\Sigma^*\times\real$ such that every set $A$ in $\CC$ has
{\em witnesses} $f$ in $\qoptsharpqp$; that is, for every $x$, if
$x\in A$ then $(x,f(x))\in A$ and otherwise $(x,f(x))\in R$. An
obvious example of such $\qoptsharpqp$-definable sets is the class
$\qma$, a quantum version of Merlin-Arthur proof systems
\cite{Kit99,Wat00,KMY01}. A less trivial example is $\nqp$, introduced
by Adleman \etal \cite{ADH97}. However, it is not yet known whether
$\pp$ is $\qoptsharpqp$-definable. We present a partial answer to this
question in Section \ref{sec:definable} by giving a new
characterization of $\pp$ in terms of quantum optimization problems.
This characterization also yields Watrous's recent result that $\qma$
is included in $\pp$ \cite{Wat01}.

The $\qoptsharpqp$-definability gives light to the relationship between
$\qoptsharpqp$ and other well-known complexity classes. {}From a
different perspective, we focus on the structure of the complexity
classes that are induced from $\qoptsharpqp$. An example of such
complexity classes is the class $\qop$. Based upon our new
characterization of $\pp$ in Section \ref{sec:definable}, we introduce
the new complexity class $\aqma$. In Section \ref{sec:QOP}, we first
show several Boolean closure properties of $\qop$ and $\aqma$. Another
important concept in complexity theory is low sets. Any set $A$ that
bears only poor information when it is used as an oracle for
relativizable class $\FF$ (that is, $\FF^A=\FF$) is called an {\em
$\FF$-low set}. We show that the class of
$\qoptsharpqp$-low sets lies between $\eqp$ and $\eqma$, which is an
error-free version of $\qma$ \cite{KMY01}. This contrasts the known
result that the class of $\sharpqp$-low sets is exactly $\eqp$
\cite{Yam99b}.
 
\section{Preliminaries}

We introduce important notions and notation in this section.

Let $\nat$, $\real$, and $\complex$ denote the sets of all natural
numbers (\eg nonnegative integers), of all real numbers, and of all
complex numbers, respectively.  Let $\nat^{+}=\nat\setminus\{0\}$.
Let $\dyadic$ be the collection of all dyadic rational numbers, where
a {\em dyadic rational number} is of the form $0.r$ or $-0.r$ for a
certain finite series $r$ of $0$s and $1$s.  In this paper, a {\em
polynomial} means a multi-variate polynomial with nonnegative integer
coefficients.

For any function $f$ and any integer $m>0$, $f^m$ denotes the function
satisfying $f^m(x)=(f(x))^m$ for all $x$.  For example, $\log^{k}n$
means $(\log n)^k$ for each $k\in\nat^{+}$.  The notation $A^{B}$,
where $A$ and $B$ are any nonempty sets, denotes the set of all total
functions mapping from $A$ to $B$: for example, $\nat^{\nat}$,
$\dyadic^{\nat}$, etc. For any two functions $f$ and $g$ with the same
domain, we say that $f$ {\em majorizes} $g$ if $f(x)\geq g(x)$ for all
$x$ in the domain of $g$.

Any element of a Hilbert space (\ie a complex vector space with the
standard inner product) of finite dimension is expressed by Dirac's
ket notation $\qubit{\phi}$. For an $n\times n$ matrix
$A=(a_{ij})_{1\leq i,j\leq n}$ over $\complex$, the notation
$A^{\dagger}$ denotes the {\em Hermitian adjoint} (\ie the transposed
complex conjugate) of $A$. Moreover, $\|A\|$ denotes the {\em operator
norm} of $A$ defined by $\sup\{\|A\qubit{\phi}\|/\|\qubit{\phi}\|\}$,
where the supremum is over all nonzero vectors $\qubit{\phi}$.  Note
that if $A$ is Hermitian, then $\|A\|=\sup_{\qubit{\phi}\neq0}
\{|\bra{\phi}A\ket{\phi}|/\|\qubit{\phi}\|\}$. A square matrix $A$ is {\em
contractive} if $\|A\|\leq 1$. See, \eg \cite{HJ85} for more detail.

\paragraph{Classical Complexity:}
Our alphabet is $\Sigma=\{0,1\}$ throughout this paper and assume the
standard (canonical and lexicographical) order on $\Sigma^*$.  The
empty string is denoted $\lambda$. For a string $x$, $|x|$ denotes the
{\em length of} $x$. We assume the reader's familiarity with
multi-tape, off-line Turing machines (TMs). We fix a {\em pairing
function} $\pair{,}$, which is a one-to-one map from
$\Sigma^*\times\Sigma^*$ to $\Sigma^*$, satisfying that, for a certain
polynomial $p$, $|\pair{x,y}|\leq p(|x|,|y|)$ for all $x$ and $y$. In
particular, we assume that the paring function preserves the length,
that is, $|\pair{x,y}|=|\pair{x',y'}|$ whenever $|x|=|x'|$ and
$|y|=|y'|$.

Let $\appcomplex$ be the set of all {\em polynomial-time approximable}
complex numbers (that is, the real and imaginary parts are
approximated to within $2^{-k}$ in time polynomial in the size of
input together with $k$). For any two sets $A$ and $B$, $A\oplus B$ is
the {\em disjoint union} of $A$ and $B$ defined by $A\oplus B=\{0x\mid
x\in A\}\cup\{1x\mid x\in B\}$.

We also assume the reader's familiarity with basic complexity classes:
$\p$, $\np$, $\co\np$, $\bpp$, $\pp$, $\pspace$, and $\dexp$. The
definition of theses complexity classes are found in, \eg
\cite{DK00,HO02}. In particular, we use the notation $\fp$ to denote 
the collection of all polynomial-time computable functions from
$\Sigma^*$ to $\Sigma^*$.  Similarly, $\fpspace$ is defined as the
collection of polynomial-space computable functions whose outputs are
also bounded by polynomials.

The class $\sharpp$ consists of all functions $f$ from $\Sigma^*$ to
$\nat$ whose values are exactly the number of accepting paths of some
polynomial-time nondeterministic TMs
\cite{Val79}. Moreover, $\gapp$ is the set of functions from
$\Sigma^*$ to $\integers$ that calculate the difference between the
number of accepting paths and the number of rejecting paths of
polynomial-time nondeterministic TMs \cite{FFK94}.  For convenience,
we translate the binary outcome of a Turing machine to an integer or a
dyadic rational number by identifying $\{0,1\}^*$ with $\integers$ or
$\{\pm 0.r\mid r\in\{0,1\}^*\}$ in the standard
order\footnote{To translate an integer or a dyadic rational number
into a string, we use the first bit (called the {\em sign bit}) of the
string to express the sign (that is, $+$ or $-$) of the number.}. By
this identification, we have the basic inclusions:
$\fp\subseteq\sharpp\subseteq\gapp\subseteq\fpspace$.

\paragraph{Quantum Complexity:}

A {\em quantum state} is a vector of unit norm in a Hilbert space.  A
{\em quantum bit} (qubit, for short) is a quantum state of
two-dimensional Hilbert space. We mainly use the standard basis
$\{\qubit{0},\qubit{1}\}$ to express a quantum state. Any quantum
state in $2^n$-dimensional Hilbert space is called a {\em quantum
string} (qustring, for short) of size $n$. The notation $\HH_{\infty}$
is used for the collection of all finite qustrings for brevity.

We use multi-tape quantum Turing machines (QTMs), defined in
\cite{BV97,ON99,Yam99a}, as a mathematical model of quantum 
computation. A multi-tape QTM is equipped with two-way infinite tapes,
tape heads, and a finite-control unit, similar to a classical TM. A
QTM follows its transition function (or algorithm), which dictates the
next move of the machine.  Formally, a $k$-tape QTM $M$ is a six-tuple
$(Q,\Sigma_1\times\Sigma_2\times \cdots \times
\Sigma_k,\Gamma_1\times\Gamma_2\times \cdots
\times\Gamma_k,q_0,Q_f,\delta)$, where $Q$ is a finite set of internal
states including the initial state $q_0$ and a set $Q_f$ of final
states, each $\Sigma_i$ is an input alphabet of tape $i$, each
$\Gamma_i$ is a tape alphabet of tape $i$ including a blank symbol and
$\Sigma_i$, and $\delta$ is a quantum transition function from
$Q\times \Gamma_1\times\cdots\times\Gamma_k$ to
$\complex^{Q\times\Gamma_1\times\cdots\times\Gamma_k\times\{L,N,R\}^k}$.
An {\em oracle QTM} is a QTM equipped with an extra query tape and two
distinguished states: a {\em pre-query} and {\em post-query}
states. Let $A$ be an oracle.  When the machine enters a pre-query
state, the string written in the query tape, say $\qubit{x}\qubit{b}$,
where $x\in\Sigma^*$ and $b\in\{0,1\}$, is changed into
$\qubit{x}\qubit{b\oplus A(x)}$ in a single step and the machine
enters a post-query state. The {\em running time} of $M$ on
multi-inputs $\vec{x}$ is the minimal number $t$ (if any) such that,
at time $t$, all computation paths of $M$ on inputs $\vec{x}$ reach
final configurations (\ie configurations with final states). We say
that $M$ on inputs $\vec{x}$ {\em halts in time $t$} if the running
time of $M$ on $\vec{x}$ is defined and is exactly $t$.

The transition function is considered as an operator that transforms a
superposition of configuration at time $t$ to another superposition of
configurations at time $t+1$. We call such an operator a {\em
time-evolution operator} (or {\em matrix}).  A QTM has {\em
$K$-amplitudes} if all the entries of its time-evolution matrix are
drawn from set $K$. A QTM is called {\em well-formed} if its
time-evolution operator is unitary (see Appendix for three local
requirements given in \cite{Yam99a}).  For simplicity, all QTMs dealt
in this paper are assumed to be well-formed unless otherwise stated. A
$k$-tape QTM $M$ is {\em stationary} if all tape heads move back to
the start cells, and $M$ is in {\em normal form} if, for every $q\in
Q_f$, there exists a series of directions $\vec{d}\in\{L,N,R\}^k$ such
that
$\delta(q,\vec{\sigma})=\qubit{q_0}\qubit{\vec{\sigma}}\qubit{\vec{d}}$
for all tape symbols
$\vec{\sigma}\in\Gamma_1\times\cdots\times\Gamma_k$.  We say that a
well-formed QTM {\em $M$ accepts input $\qubit{\phi}$ with probability
$\alpha$} if $M$ halts in a final configuration in which, when
observed, bit $1$ is found in the start cell of the output tape with
probability $\alpha$. In this case, we also say that $M$ {\em rejects
input $\qubit{\phi}$ with probability $1-\alpha$}.

A function $f$ from $\Sigma^*$ to the unit real interval $[0,1]$ is in
$\sharpqp$ (``sharp'' QP) if there exists a polynomial-time
well-formed QTM $M$ with $\appcomplex$-amplitudes such that, for every
$x$, $f(x)$ is the probability that $M$ accepts $x$ \cite{Yam99b}.  In
this case, we simply say that $M$ {\em witnesses $f$}. A function $f$
from $\Sigma^*$ to $[-1,1]$ is in $\gapqp$ if there exists a
polynomial-time well-formed QTM $M$ with $\appcomplex$-amplitudes such
that, for every $x$, $f(x)$ is the difference between the acceptance
probability of $M$ on input $x$ and the rejection probability of $M$
on $x$ \cite{Yam99b}.  Let $\feqp$ be the collection of all functions
from $\Sigma^*$ to $\Sigma^*$ whose outputs are produced by
polynomial-time, $\appcomplex$-amplitude, well-formed QTMs with
certainty \cite{Yam99b,BD99}.  A set $A$ is in $\bqp$ if there exists a
function $f$ in $\sharpqp$ such that, for every $x$, if $x\in A$ then
$f(x)\geq 3/4$ and otherwise $f(x)\leq 1/4$ \cite{BV97}.  A set $A$ is
in $\nqp$ if there exists a function $f$ in $\sharpqp$ such that, for
every $x$, $x\in A$ iff $f(x)>0$ \cite{ADH97}.
\ms

\n{\sf Remark on the convention of quantum extension:} Although we 
normally deal with classical inputs for a QTM $M$, we sometimes feed
$M$ with quantum states. In this way, however, we can naturally expand
the definition of a function $f$ based on classical strings to a
function $\check{f}$ based on qustrings without altering $M$ that
defines $f$. Such $\check{f}$ is called the {\em quantum extension} of
$f$. By abusing the notation, we use the same symbol $f$ to cope with
both functions. We use the same convention for a set of strings.
\ms

Using the aforementioned convention of quantum extension, we define
$\qma$ as follows: a set $A$ is in $\qma$ if there exist a polynomial
$p$ and a function $f$ in $\sharpqp$ such that, for every $x$, if
$x\in A$ then $f(\qubit{x}\qubit{\phi})\geq 3/4$ for a certain
qustring $\qubit{\phi}$ of size $p(|x|)$ and if $x\not\in A$ then
$f(\qubit{x}\qubit{\phi})\leq 1/4$ for every qustring $\qubit{\phi}$
of size $p(|x|)$ \cite{Kit99,Wat00}\footnote{Kitaev
\cite{Kit99} and Watrous \cite{Wat00} defined the class $\qma$ based
on the quantum circuit model. In this paper, for our convenience, we
use the quantum Turing machine model.}.

\ignore{
\paragraph{Classical Sets vs. Quantum Sets.}
When we discuss quantum complexity classes, it appears convenient to
consider a set of qustrings instead of a set of strings. We naturally
expand a set $A$ in a quantum complexity class, such as $\bqp$ and
$\qma$, to a subset of $\HH_{\infty}$ by considering a qustring as an
input to a basis quantum Turing machine. For example, a subset $A$ of
$\HH_{\infty}$ is in $\bqp$ if there exist a constant $\epsilon>0$ and
a polynomial-time quantum Turing machine $M$ such that, for every
qustring $\qubit{\phi}$ of size $n$, $M$ accepts $\qubit{\phi}$ with
probability at least $\frac{1}{2}+\epsilon$ if $\qubit{\phi}\in A$;
otherwise, $M$ rejects $\qubit{\phi}$ with probability at least
$\frac{1}{2}+\epsilon$. To distinguish such a set from a classical
set, we call a set of qustrings a {\em quantum set} for brevity. A
collection of quantum sets is called a {\em quantum (complexity)
class}. We write $\widetilde{C}$ to denote such a class.

Due to quantum interference, $\widetilde{\CC}$ may not enjoy the same
structural property as $\CC$ does. For example, $\bqp$ enjoys the
amplification property; that is, we can boost its accepting
probability from $\frac{1}{2}+\epsilon$ to $1-2^{-p(n)}$, where $p$ is
any polynomial. However, it is not known if $\widetilde{\bqp}$ enjoys
the same property.  }

\section{Krentel's Framework for Optimization Problems}

Optimization problems have arisen in many areas of computer
science. Most $\np$-complete decision problems, for instance,
naturally yield their optimization counterparts. In the 1980s, Krentel
\cite{Kren88} made a systematic approach toward NP optimization problems.
In particular, he studied the problems of finding the maximal (as well
as minimal) outcome of a polynomial-time nondeterministic computation
and he introduced the function class, called $\optp$, that constitutes
all such optimization problems.

\begin{definition}\label{def:OptP}{\rm \cite{Kren88}}\hs{2}
A function from $\Sigma^*$ to $\nat$ is in $\optp$ if there exists a
polynomial-time nondeterministic TM $M$ such that, for every $x$, (i)
every computation path of $M$ on input $x$ terminates with binary
strings (which are interpreted as natural numbers) on its output tape
and accepts and (ii) $f(x)$ is the maximum\footnote{Although Krentel
also includes the minimization problems into $\optp$, this paper
focuses only on maximization problems as in \cite{KST89} and
Definition \ref{def:OptP} does not include any minimization problems.}
output value of $M$ on input $x$.
\end{definition}

Using the notion of $\fp$-functions, we rephrase Definition
\ref{def:OptP} in the following way: a function $f$ is in $\optp$ iff
there exist a polynomial $p$ and a function $g\in\fp$ such that, for
every $x$,
\[
 f(x)=\max\{g_{s}(x)\mid s \mbox{ is any string of length $p(|x|)$ }\},
\]
where $g_{s}(x)$ means $g(\pair{x,s})$ for each $s$ viewed as an
index.  This characterization enables us to generalize $\optp$ to
$\mathrm{\bf Opt}\FF$ by replacing an $\fp$-function $g$ with another
function from a more general function class $\FF$.

\begin{definition}
Let $\FF$ be any set of functions from $\Sigma^*$ to $\real$. A
function $f$ from $\Sigma^*$ to $\real$ is in $\mathrm{\bf Opt}\FF$ if
there exists a polynomial $p$ and a function $g$ in $\FF$ such that,
for every $x$, $f(x)=\max\{g_{s}(x)\mid s\in\Sigma^{p(|x|)}\}$, where
$g_{s}(x)=g(\pair{x,s})$. The subscript $s$ is called a {\em classical
index}. The class $\FF$ is called the {\em underlying class} of
$\mathrm{\bf Opt}\FF$.
\end{definition}

With this general notation, $\mathrm{\bf Opt}\fp$ coincides with
$\optp$.  Another example obtained from the above definition is the
function class $\optsharpp$ by taking $\sharpp$ as an underlying
class.  Clearly, $\optp\cup\sharpp \subseteq\optsharpp$ since
$\fp\subseteq\sharpp$ (by identifying binary strings with natural
numbers).  Moreover, we can define the class $\optsharpqp$ by taking
$\sharpqp$ as an underlying set.  This class naturally expands
$\optsharpp$ but stays within $\sharpqp^{\np^{\pp}}$.
 
\begin{lemma}\label{lemma:opt-inclusion}
\begin{enumerate}
\item For every $f\in\optsharpp$, there exist two functions
$g\in\optsharpqp$ and $\ell\in\fp$ such that, for all $x$,
$f(x)=g(x)\ell(1^{|x|})$.
\vs{-2}
\item $\optsharpqp\subseteq\sharpqp^{\np^{\pp}}$.
\end{enumerate}
\end{lemma}

For the proof, we need the following result from a revision of
\cite{Yam99b}.

\begin{lemma}\label{lemma:pp-sharpqp}{\rm \cite[revision]{Yam99b}}\hs{1}
 Let $A$ be any set. The following three statements are all
 equivalent: (i) $A\in\pp$; (ii) there exist two functions
 $f,g\in\sharpqp$ such that, for every $x$, $x\in A$ iff $f(x)>g(x)$;
 and (iii) there exist two functions $f,g\in\gapqp$ such that, for
 every $x$, $x\in A$ iff $f(x)>g(x)$.
\end{lemma}

\begin{proofof}{Lemma \ref{lemma:opt-inclusion}}
1) Let $f$ be any function in $\optsharpp$. That is, there exists a
polynomial $p$ and a function $h\in\sharpp$ such that
$f(x)=\max\{h_{s}(x)\mid s\in\Sigma^{p(|x|)}\}$, where
$h_{s}(x)=h(\pair{x,s})$.  By the argument used in \cite{Yam99b}, we
can define $k\in\sharpqp$ and $\ell\in\fp$ that satisfy the following
condition: $k(\pair{x,s})=0$ if $|s|\neq p(|x|)$, and otherwise,
$k(\pair{x,s})\ell(1^{|x|})=h(\pair{x,s})$. For the desired $g$,
define $g(x)=\max\{k_{s}(x)\mid s\in\Sigma^{p(|x|)}\}$ for each $x$.

2) Let $f$ be any function in $\optsharpqp$ and take a $g\in\sharpqp$
and a polynomial $p$ such that $f(x)=\max\{g_{s}(x)\mid
s\in\Sigma^{p(|x|)}\}$ for all $x$. Choose the lexicographically
minimal string $s_{x}$ such that $f(x)= g(\pair{x,s_{x}})$ and
$|s_{x}|=p(|x|)$. Define $A=\{\pair{x,s,t,y}\mid \exists z(s\leq z\leq
t\wedge g_{z}(x)\geq 0.y)\}$. Note that $A$ belongs to $\np^{\pp}$ by
Lemma
\ref{lemma:pp-sharpqp}. The function $f$ is computed as follows using
$A$ as an oracle. Let $x$ be any input of length $n$. By binary
search, we find a string $\hat{y}$ such that (i) $g_{s_{x}}(x)\geq
0.\hat{y}$ and (ii) for every $z\in\Sigma^{p(|x|)}$,
$g_{z}(x)<g_{s_{x}}(x)$ implies $g_{z}(x)<0.\hat{y}$. Then, by
decreasing the interval $[s,t]$ in a way similar to the binary search,
we can find a string $s_{0}$ of length $p(|x|)$ such that
$g_{s_{0}}(x)\geq 0.\hat{y}$. At last, we compute $g_{s_{0}}(x)$ in a
quantum fashion. Thus, $f$ is in
$\sharpqp^{A}\subseteq\sharpqp^{\np^{\pp}}$.
\end{proofof}

We next show that the collapse $\optsharpqp=\sharpqp$ is unlikely. To
state Proposition \ref{prop:EQP-WQP}, we recall from \cite{Yam99b} the
class $\wqp$, which is a quantum analogue of $\wpp$ in \cite{FFK94}.

\begin{definition}{\rm \cite{Yam99b}}\hs{1}
A set $S$ is in $\wqp$ (``wide'' QP) if there exist two functions
$f\in\sharpqp$ and $g\in\feqp$ such that, for every $x$, (i) $g(x)\in
(0,1]\cap\rational$ and (ii) if $x\in S$ then $f(x)= g(x)$ and
otherwise $f(x)=0$, where we identify a string with a rational number
expressed as a pair of integers (\eg $g(x)=\frac{1}{3}$).
\end{definition}

Note that $\eqp\subseteq\wqp\subseteq\nqp\subseteq\pp$. However, we do
not know any relationship between $\wqp$ and $\bqp$.

\begin{proposition}\label{prop:EQP-WQP}
$\pp=\eqp$ $\Longrightarrow$ $\sharpqp=\optsharpqp$ $\Longrightarrow$
$\pp=\wqp$.
\end{proposition}

\begin{proof}
If $\eqp=\pp$ then
$\np^{\pp}\subseteq\pp^{\pp}\subseteq\eqp^{\eqp}=\eqp$. Since
$\sharpqp^{\eqp}=\sharpqp$
\cite{Yam99b}, by Lemma \ref{lemma:opt-inclusion}, we obtain 
$\optsharpqp\subseteq\sharpqp$.

For the second implication, assume that $\sharpqp=\optsharpqp$. Let
$A$ be an arbitrary set in $\pp$. As shown in \cite{Yam99b}, $\pp$
coincides with its quantum analogue $\pqp_{\appcomplex}$. Thus, there
exists a function $f\in\sharpqp$ such that, for every $x$, (i) if
$x\in A$ then $f(x)>1/2$ and (ii) if $x\not\in A$ then $f(x)<
1/2$. Define $g(x)=\max\{f(x),\frac{1}{2}\}$ for every $x$. This $g$
is clearly in $\optsharpqp$ and thus in $\sharpqp$. By the definition
of $g$, if $x\in A$ then $g(x)>1/2$ and otherwise $g(x)=1/2$. Next,
define $h(x)=g(x)-\frac{1}{2}$ for all $x$'s. Since
$\gapqp=\sharpqp-\sharpqp$ \cite{Yam99b}, $h$ is in $\gapqp$. Thus,
$h^2$ is in $\sharpqp$ \cite{Yam99b}. For this $h^2$, it follows that
$x\in A$ implies $h^2(x)>0$ and $x\not\in A$ implies $h^2(x)=0$. At
this moment, we obtain that $A\in\nqp$.

Since $h^2$ is in $\sharpqp$, there exists a polynomial $p$ such that,
for every $x$, if $h^2(x)>0$ then $h^2(x)>2^{-p(|x|)}$ because the
acceptance probability of a QTM is expressed in terms of a polynomial
in the transition amplitudes from $\appcomplex$. To complete the
proof, we define $k(x)=\min\{h^2(x),2^{-p(|x|)}\}$ for every $x$. It
follows that $x\in A$ implies $k(x)=2^{-p(|x|)}$ and $x\not\in A$
implies $k(x)=0$. To see that $k$ is in $\sharpqp$, consider the fact
that $1-k(x)=\max\{1-h^2(x),1-2^{-p(|x|)}\}$ is in $\optsharpqp$,
which is $\sharpqp$ by our assumption. Thus, $1-(1-k(x))$ is also in
$\sharpqp$, which implies that $k$ is in $\sharpqp$.
\end{proof}

At the end of this section, we note that Krentel's framework has been
extended in several different manners in the literature
\cite{CT91,BKT98}.

\section{Quantum Optimization Problems}\label{sec:quantum-opt}

Krentel's optimization problems are to maximize the value of indexed
functions chosen from underlying class $\FF$. As shown in the previous
section, Krentel's framework can cope with the class
$\optsharpqp$. However, $\optsharpqp$ is a concoction of a classical
indexing system and quantum computations. In this section, we truly
expand Krentel's framework and introduce quantum optimization
problems. Our quantum optimization problem uses a quantum computation
together with a quantum index, which is a qustring of polynomial
size. We begin with the general definition, paving a road to our study
of $\qoptsharpqp$.

\begin{definition}\label{def:opt-qp}
Let $\FF$ be a set of functions from $\Sigma^*\times\HH_{\infty}$ to
$\real$. A quantum optimization problem $f$ from $\Sigma^*$ to $\real$
is in $\mathrm{\bf Qopt}\FF$ if there exist a polynomial $p$ and a
function $g\in \FF$ such that, for all $x$,
\[
f(x)=\sup\{g_{\qubit{\phi}}(x)\mid
\mbox{ $\qubit{\phi}$ is any qustring of size $p(|x|)$ }\},
\]
where $g_{\qubit{\phi}}(x)=g(x,\qubit{\phi})$. The subscript
$\qubit{\phi}$ is called a {\em quantum index}. For simplicity, we say
that $g$ {\em witnesses} $f$.
\end{definition}

In the course of a study on a quantum optimization problem, it is
rather convenient to optimize the acceptance probability of a
polynomial-time well-formed QTM indexed with a quantum state. {}From
this reason, we mainly study the class $\qoptsharpqp$ throughout this
paper and leave other platforms to the interested reader.  This class
$\qoptsharpqp$ consists of all quantum optimization problems $f$ such
that there exist a polynomial $p$ and a multi-tape, polynomial-time,
well-formed QTM $M$ with $\appcomplex$-amplitudes satisfying that, for
all $x$, $f(x)$ is equal to the supremum, over all qustring
$\qubit{\phi}$ of size $p(|x|)$, of the probability that $M$ accepts
$(\qubit{x},\qubit{\phi})$. The class $\qoptsharpqp$ naturally
includes $\optsharpqp$.
\ms

\n{\sf Observation:} The size factor $p$ of a quantum index 
$\qubit{\phi}$ in Definition \ref{def:opt-qp} can be replaced by any
polynomial $q$ that majorizes $p$. This is shown easily by ignoring
the extra $q(|x|)-p(|x|)$ qubits on input $x$ because those qubits
that are not accessed by a QTM do not affect the acceptance
probability of the QTM.
\ms

Solving a quantum optimization problem is closely related to finding
the maximal eigenvalue of a certain type of positive semidefinite,
contractive, Hermitian matrix.  In what follows, we clarify this
relationship.  Assume that we have a multi-tape well-formed QTM
witnessing a quantum optimization problem $f$. Let $p$ be a polynomial
expressing the size of a quantum index. Without loss of generality, we
can assume that $M$ is stationary in normal form
\cite{BV97,Yam99a}. As shown in \cite{Yam99a}, there exists a
reversing QTM for $M$, denoted $M^{\dagger}$. We then introduce the
new QTM, called $N_{M,p}$, that behaves as follows:
\begin{quote}
{\sf On input $(x,t)$ given in the input tape, if $|t|\neq p(|x|)$
then skip the following procedure and halt. Let $n=|x|$ and assume
that $t\in\Sigma^{p(n)}$. Copy $x$ onto a new tape, called the storage
tape. Run $M$ on input $(x,t)$. When it halts in polynomial time, copy
$M$'s output bit (either $\qubit{0}$ or $\qubit{1}$) onto another new
tape, called the result tape. Run the reversing machine $M^{\dagger}$
on the final configuration excluding the storage tape and the result
tape. After $M^{\dagger}$ halts in polynomial time, if $N_{M,p}$'s
configuration consists only of $(x,s)$ in the input tape, where $s$ is
a certain string of length $p(|x|)$, of $x$ in the storage tape, of
$1$ in the result tape (and empty elsewhere), then move $(x,s)$ into
the output tape and empty all the other tapes and halt. Otherwise,
write the blank symbol in the output tape and halts.}
\end{quote}
By an appropriate implementation, we can make all the computation
paths of $N_{M,p}$ on each input terminate simultaneously.  Now, fix
$x$ and let $n=|x|$. For each pair
$(s,t)\in\Sigma^{p(n)}\times\Sigma^{p(n)}$, define $\alpha_{x,s,t}$ to
be the amplitude of a unique final configuration in which, starting
with input $(x,t)$, $N_{M,p}$ outputs $(x,s)$ in a unique final inner
state.  Let $P_{M,p,x}$ be the $2^{p(n)}\times2^{p(n)}$ matrix
$(\alpha_{x,s,t})_{s,t\in\Sigma^{p(n)}}$. It is not difficult to show
by the definition of $N_{M,p}$ that $P_{M,p,x}$ is Hermitian, positive
semidefinite, and contractive.  Thus, each $\alpha_{x,s,t}$ is a real
number and $\alpha_{x,s,t}=\alpha_{x,t,s}$ for all pairs
$(s,t)\in\Sigma^{p(n)}\times\Sigma^{p(n)}$.  Note that the acceptance
probability of $M$ on input $(x,\qubit{\phi})$ is exactly
$|\bra{\phi}P_{M,p,x}\ket{\phi}|$. Thus,
\begin{eqnarray*}
f(x) &=& \sup\{|\bra{\phi}P_{M,p,x}\ket{\phi}|
\mid \mbox{ $\qubit{\phi}$ is any qustring of size $p(n)$ }\} \\ 
&=&
\|P_{M,p,x}\|,
\end{eqnarray*} 
which coincides with the maximal eigenvalue $\lambda$ of $P_{M,p,x}$
(see, \eg \cite{HJ85}). A similar construction has been shown in the
literature \cite{BBBV97,Yam99a}.

In a quantum setting, ``squaring'' becomes a unique operation because
of the reversibility nature of quantum computation. For example, as
shown in \cite{Yam99b}, $f\in\gapqp$ implies $f^2\in\sharpqp$. In the
next lemma, we show that squaring does not increase the size of a
quantum index.

\begin{lemma}\label{lemma:squaring}
Let $p$ be any polynomial, $g$ any function in $\sharpqp$, and $f$ any
function in $\qoptsharpqp$. Assume that
$f(x)=\sup\{g(x,\qubit{\phi})\mid \mbox{ $\qubit{\phi}$ is any
qustring of size $p(|x|)$ }\}$ for all $x$.  Then, there exists a
function $h$ in $\sharpqp$ such that
$f^2(x)=\sup\{h(x,\qubit{\phi})\mid \mbox{ $\qubit{\phi}$ is any
qustring of size $p(|x|)$ }\}$ for all $x$.
\end{lemma}

\begin{proof}
Since $g\in\sharpqp$, let $M$ be a multi-tape,
$\appcomplex$-amplitude, stationary, well-formed QTM $M$ in normal
form that witnesses $g$. By the aforementioned argument, $f(x)$ equals
$\|P_{M,p,x}\|$. Since $P_{M,p}$ is Hermitian,
$f^2(x)=\|P_{M,p,x}\|^2=\|P_{M,p}^2\|$. We slightly modify the QTM
$N_{M,p}$ and define the new QTM, called $N$, in the following
fashion:
\begin{quote}
{\sf On input $(x,t)$, if $|t|\neq p(|x|)$ then reject the input. Let
$n=|x|$ and assume that $t\in\Sigma^{p(n)}$.  Run $N_{M,p}$ on input
$(x,t)$. If $N_{M,p}$ outputs $(x,s)$ for a certain string $s$ of
length $p(|x|)$, then accept the input. Otherwise, reject the input.}
\end{quote}
For the desired $h$, define $h(x,t)$ to be the acceptance probability
of $N$ on input $(x,t)$. Note that $N_{M,p}$ accepts input
$(x,\qubit{\phi})$ with probability exactly
$|\bra{\phi}P_{M,p,x}^{\dagger}P_{M,p,x}\ket{\phi}|$, which equals
$|\bra{\phi}P_{M,p,x}^2\ket{\phi}|$. Hence,
\begin{eqnarray*}
\|P_{M,p,x}^2\| 
&=& \sup\{|\bra{\phi}P_{M,p,x}^2\ket{\phi}| \mid \mbox{ $\qubit{\phi}$
is any qustring of size $p(|x|)$ }\} \\ 
&=& \sup\{h(x,\qubit{\phi})\mid 
\mbox{ $\qubit{\phi}$ is any qustring of size $p(|x|)$ }\}.
\end{eqnarray*}
Since $f^2(x)=\|P_{M,p,x}^2\|$, we obtain the desired equation. 
\end{proof}

\section{Fundamental Properties of Qopt\#QP}\label{sec:properties}

We examine the fundamental properties of quantum optimization problems
in $\qoptsharpqp$. In what follows, we show that $\qoptsharpqp$ enjoys
important closure properties, such as multiplication, exponentiation,
limited addition, and limited composition.

Firstly, we show that $\qoptsharpqp$ is closed under composition with
$\fp$-functions. To be more precise, for any two function classes
$\FF$ and $\GG$, let $\FF\circ
\GG$ denote the class of functions $f\circ g$, where $f\in\FF$ and
$g\in\GG$ and $f\circ g$ is the composition defined as $f\circ
g(x)=f(g(x))$ for all $x$ in the domain of $g$. Using this notation,
our claim is expressed as follows. The proof of the claim is
immediate.

\begin{lemma}
$\qoptsharpqp\circ\fp = \qoptsharpqp$.
\end{lemma}

In the previous section, we have shown that solving a quantum
optimization problem is equivalent to finding the maximal eigenvalue
of a certain positive semidefinite, contract, Hermitian matrix. We use
this characterization to show the fundamental properties of quantum
optimization problems. To describe the claim, we need the notion of an
$\ell$-qubit source.

A {\em qubit ensemble} is a sequence of qustrings with index set $I$.
A qubit ensemble $\{\qubit{\phi_x}\}_{x\in I}$ is called an {\em
$\ell$-qubit source} if each $\qubit{\phi_x}$ is a qustring of size
$\ell(|x|)$ and there exists a polynomial-time,
$\appcomplex$-amplitude, well-formed, clean QTM that generates
$\qubit{\phi_x}$ on input $x$, where a QTM is called {\em clean} if it
is stationary in normal form and all the tapes except the output tape
are empty when it halts with one distinguished final state.

\begin{lemma}\label{lemma:property-1}
Let $q$ be a polynomial. Let $\ell$ be any function in $\nat^{\nat}$
such that $\ell(n)\in O(\log n)$ and let $\{\qubit{\phi_x}\}_{x\in
\Sigma^*}$ be any $\ell$-qubit source.  Assume that $f$ is a function
in $\qoptsharpqp$.  The following functions all belong to
$\qoptsharpqp$.
\begin{enumerate}
\item $g(x)=\max\{f(\pair{x,y})\mid y\in\Sigma^{q(|x|)}\}$.
\vs{-2}
\item $g(x)=\sum_{y:|y|=\ell(|x|)}|\measure{y}{\phi_x}|^2\cdot
f(\pair{x,y})$.
\vs{-2}
\item $g(x)=\prod_{y:|y|=\ell(|x|)}f(\pair{x,y})$.
\vs{-2}
\item $g(\pair{x,y})=f(x)^{|y|}$.
\end{enumerate}
\end{lemma}

\begin{proof}
1) Take a polynomial $p$ and a function $h\in\sharpqp$ such that
$f(\pair{x,y})=\sup\{h(x,y,\qubit{\phi})\mid
\mbox{ $\qubit{\phi}$ is any qustring of size $p(|x|)$ }\}$ for 
any strings $x$ and $y\in\Sigma^{q(|x|)}$. Since $h\in\sharpqp$, there
exists an appropriate QTM $M$ that witnesses $h$.  For simplicity,
assume that there exists a polynomial $r$ such that
$|\pair{x,y}|=r(|x|)$ for all $y\in\Sigma^{q(|x|)}$.

Define the new QTM $N$ as follow. Let $(x,\qubit{\psi})$ be any
quantum input, where $n=|x|$ and $\qubit{\psi}$ is any qustring of
size $q(n)+p(n)$. Note that $\qubit{\psi}$ is written in the form
$\sum_{y:|y|=q(n)}\beta_y\qubit{y}\otimes\qubit{\phi_y}$ for a certain
series of qustrings $\{\qubit{\phi_y}\}_{y\in\Sigma^{q(n)}}$ and a
certain series of complex numbers $\{\beta_y\}_{y\in\Sigma^{q(n)}}$
satisfying $\sum_{y:|y|=q(n)}|\beta_y|^2=1$.
\begin{quote}
{\sf First, observe the first $q(n)$ qubits (\ie $\qubit{y}$) of the
tape content $\qubit{\psi}$ and copy the result onto a new tape to
remember it.  After the observation, we obtain $y$ and a quantum state
$\beta_y\qubit{\phi_y}$. Then, simulate $M$ on inputs $(x,y)$ and
$\beta_y\qubit{\phi_y}$.}
\end{quote}
Let $k(x,\qubit{\psi})$ denote the acceptance probability of $N$ on
input $(x,\qubit{\psi})$. Obviously, $k(x,\qubit{\psi})=
\sum_{y:|y|=q(n)}|\beta_y|^2 h(x,y,\qubit{\phi_y})$.

Consider the supremum $\sup_{\qubit{\psi}}\{k(x,\qubit{\psi})\}$,
where $\qubit{\psi}$ runs over all qustrings of size $q(n)+p(n)$. By a
simple calculation, $\sup_{\qubit{\psi}}\{k(x,\qubit{\psi})\}=
\sup_{\{\beta_y\}_{y},\qubit{\phi_y}}
\{\sum_{y:|y|=q(n)}|\beta_y|^{2}h(x,y,\qubit{\phi_y})\}$, which is 
equal to $\sup_{\{\beta_y\}_{y}}\{\sum_{y:|y|=q(n)}|\beta_y|^2
f(\pair{x,y})\}$, where $\{\beta_{y}\}_{y\in\Sigma^{q(|x|)}}$ runs
over all sequences of complex numbers with
$\sum_{y:|y|=q(|x|)}|\beta_y|^2=1$ and each $\qubit{\phi_y}$ runs all
qustrings of size $q(n)$.  By the property of convex combination with
$\sum_{y:|y|=q(n)}|\beta_y|^2=1$, the value
$\sup_{\{\beta_y\}_{y}}\{\sum_{y:|y|=q(n)}|\beta_y|^2 f(\pair{x,y})\}$
equals $\max\{f(\pair{x,y})\mid y\in\Sigma^{q(n)}\}$. Therefore, $N$
witnesses $g$.

2) For simplicity, we assume the existence of a polynomial $r$ such
that $2^{\ell(n)}\leq r(n)$ for all $n\in\nat$.  Without loss of
generality, we can assume that there exist a polynomial $p$ and a
function $h\in\sharpqp$ which, for any strings $x$ and
$y\in\Sigma^{\ell(|x|)}$, force $\sup\{h(x,y,\qubit{\phi})\mid
\mbox{ $\qubit{\phi}$ is any qustring of size $p(|x|)$ }\}$ to be 
$f(\pair{x,y})$. Let $M$ be an appropriate QTM $M$ that witnesses $h$.

Consider the following QTM $N$. Let $x$ (say, $n=|x|$) be any input
string and $\qubit{\psi}$ any input qustring of size $r(n)p(n)$. 
We section $\qubit{\psi}$ into $r(n)$ blocks of equal size $p(n)$. The
first $2^{\ell(n)}$ blocks are assumed to be indexed with strings of
length $\ell(n)$. 
\begin{quote}
{\sf Generate qustring $\qubit{\phi_{x}}$ in a new tape and then
observe the tape content. Let $y$ be the result after the
observation. Now, copy $y$ onto a new tape and then simulate $M$ on
input $(x,y)$ as well as the content of the $y$th block of
$\qubit{\psi}$.}
\end{quote}

For each $y$, let $\qubit{\psi_y}$ be any qustring of size $p(n)$ that
achieves the maximal acceptance probability of $M$ on input
$(x,y,\qubit{\psi_y})$; that is
$f(\pair{x,y})=h(x,y,\qubit{\psi_y})$. Note that, for any qustring
$\qubit{\psi}$ of size at least $p(n)$, the acceptance probability of
$M$ on input $(x,y)$ accessing only the first $p(n)$ qubits of
$\qubit{\psi}$ cannot be more than $f(\pair{x,y})$.  Hence, to
maximize the acceptance probability of $N$, $\qubit{\psi}$ must be the
tensor product $(\bigotimes_{y:|y|=\ell(n)}\qubit{\psi_y})\otimes
\qubit{\xi}$, where $\qubit{\xi}$ is any qustring of size
$(r(n)-2^{\ell(n)})p(n)$.  In this case, the acceptance probability of
$N$ on input $(x,\qubit{\psi})$, after $y$ is observed, is
$|\measure{y}{\phi_{x}}|^2h(x,y,\qubit{\psi_y})$. Note that any two
simulations on different $y$'s do not interfere each other. Hence, $N$
witnesses $g$.

3) Following 2), we assume that $2^{\ell(n)}\leq r(n)$ for all
$n\in\nat$ and $f(\pair{x,y})=\sup\{h(x,y,\qubit{\phi})\mid
\mbox{ $\qubit{\phi}$ is any qustring of size $p(|x|)$ }\}$ for any 
strings $x$ and $y\in\Sigma^{\ell(|x|)}$. This $h$ is witnessed by a
QTM $M$.  The new QTM $N$ works as follows. Let $x$ be any string of
length $n$ and $\qubit{\psi}$ any qustring of size $r(n)p(n)$. We use
the same sectioning of $\qubit{\psi}$ as in 2).
\begin{quote}
{\sf Set $y=0^{\ell(n)}$ initially. By incrementing $y$
lexicographically until $1^{\ell(n)}$, simulate recursively $M$ on
inputs $\pair{x,y}$ together with the content of the $y$th block of
$\qubit{\psi}$. Before each simulation, copy the value $y$ onto a
separate tape to make each round of simulation independent. If $M$
reaches a rejecting configuration at a certain round, then reject the
input. Otherwise, accept the input.}
\end{quote}

Similar to 2), when $\qubit{\psi}$ is the tensor product of the form
$(\bigotimes_{y:|y|=\ell(n)}\qubit{\psi_y})\otimes \qubit{\xi}$, we
can achieve the maximal acceptance probability of $N$ on input
$(x,\qubit{\psi})$. Thus, $N$ witnesses $g$.
 
4) This is basically an extension of the proof of Lemma
\ref{lemma:squaring}. Since $f\in\qoptsharpqp$, let $p$ be a
polynomial and $M$ be a QTM that witnesses $f$. Let $N_{M,p}$ be the
QTM induced from $M$ as described in Section
\ref{sec:quantum-opt}. For our proof, we slightly modify $N_{M,p}$ in
such way that, instead of moving $(x,s)$ from the input tape to the
output tape, it keeps $(x,s)$ in the input tape and erases only the
storage-tape content (keeping 1 in the result tape).

Now, consider the following QTM $P$. Let $x$ be any input string of
length $n$ and $y$ be any string of length $m$. Let $m=2k+j$, where
$k\in\nat$ and $j\in\{0,1\}$. We prepare a new tape, called the {\em
checking tape}.
\begin{quote}
{\sf On input $(x,y,\qubit{\psi})$ given in tape 1, repeat the
following procedure (*) by incrementing $i$ by one from $1$ to
$k$. For convenience, we call the initial configuration round 0.

\hs{5}(*) At round $i$, $P$ simulate $N_{M,p}$ starting with the
configuration left from the previous round, and when it halts, move
the one-bit content of the result tape into the $i$th cell of the
checking tape.

After this procedure, when $j=1$, we make one additional round:
simulate $M$ starting with the configuration left from the previous
round and, when it halts, write the outcome (either $0$ or $1$) into
the $k+j$th cell of the checking tape. Finally, observe the checking
tape and accept the input iff the $k+j$ cells consists only of $1$s.}
\end{quote}

Let $\qubit{\psi_x}$ be any quantum index that maximizes the
acceptance probability of $M$ on input $(x,\qubit{\psi_x})$. From the
proof of Lemma \ref{lemma:squaring}, $N_{M,p}$ accepts
$(x,\qubit{\phi_x})$ with probability $f^2(x)$. More generally, we can
show by induction that, after each round $i$, if we observe the
checking tape with observable $\qubit{1^i}$ then we obtain the quantum
state $f^{2i}(x)\qubit{x}\qubit{\phi_x}$ in the input tape. In case
where $j=1$, the additional round contributes to a multiplicative
factor of $f(x)$.  Therefore, $P$ accepts $(x,\qubit{\psi_x})$ with
probability exactly $f^{2k+j}(x)$, which is $f^{m}(x)$.

It is also important to note that the size of quantum index is
independent of $y$.
\end{proof}

The class $\qoptsharpqp$ is shown to be robust in the following sense.

\begin{proposition}\label{prop:robust}
$\mathrm{\bf Qopt}(\optsharpqp) = \mathrm{\bf Opt}(\qoptsharpqp) =
\qoptsharpqp$. 
\end{proposition}

As for the definition of $\mathrm{\bf Qopt}(\optsharpqp)$, it is
important to note that, by our convention of quantum extension, $f$ is
in $\mathrm{\bf Qopt}(\optsharpqp)$ iff there exists a function
$h\in\sharpqp$ such that, for every $x$,
$f(x)=\sup_{\qubit{\phi}}\max_{s}\{h(x,\qubit{\phi},s)\}$, where
$\qubit{\phi}$ runs over all qustrings of polynomial size and $s$ runs
over all strings of polynomial length.

\begin{proofof}{Proposition \ref{prop:robust}}\hs{1}
It follows from $\sharpqp\subseteq\optsharpqp$ that
$\qoptsharpqp\subseteq\mathrm{\bf Qopt}(\optsharpqp)$. We then show
that $\mathrm{\bf Qopt}(\optsharpqp)\subseteq\mathrm{\bf
Opt}(\qoptsharpqp)$. This follows from the fact that we can swap the
``max'' operator and the ``sup'' operator.  The inclusion $\mathrm{\bf
Opt}(\qoptsharpqp) \subseteq\qoptsharpqp$ follows from Lemma
\ref{lemma:property-1}(1).
\end{proofof}

Proposition \ref{prop:robust} seems unlikely to be extended to
$\mathrm{\bf Qopt}(\qoptsharpqp) = \qoptsharpqp$ because of the
inability to distinguish a tensor product of two short 
quantum indices from one large quantum index. We
conjecture that $\mathrm{\bf Qopt}(\qoptsharpqp) \neq
\qoptsharpqp$.

This also suggests a generalization of $\qoptsharpqp$ into
$\mathrm{\bf Qopt}_{k}\sharpqp$ for each $k\in\nat^{+}$ by taking a
tensor product of $k$ quantum indices: a function $f$ is in
$\mathrm{\bf Qopt}_{k}\sharpqp$ if there exists a polynomial $p$ and a
function $g\in\sharpqp$ such that, for every $x$,
$f(x)=\sup\{g(x,\qubit{\phi_1}\otimes\qubit{\phi_2}
\otimes\cdots\otimes\qubit{\phi_k})\}$, where the supremum is over all
$k$-tuples $(\qubit{\phi_1}, \qubit{\phi_2},\ldots,\qubit{\phi_k})$ of
strings of size $p(|x|)$. With this notation, we can show that
$\mathrm{\bf Qopt}(\qoptsharpqp) = \mathrm{\bf Qopt}_{2}\sharpqp$.

\section{Approximate Reduction and Universality}\label{sec:universal}

Based upon his metric reducibility between NP optimization problems,
Krentel \cite{Kren88} introduced the notion of {\em complete problems}
for $\optp$. There are also many other reductions used for NP
optimization problems in the literature. Krentel's complete problems
constitute the hardest problems in $\optp$.  It is natural to consider
a similar notion among quantum optimization problems. However,
problems in $\qoptsharpqp$ are functions computed by well-formed
$\appcomplex$-amplitude QTMs and thus, there is no single QTM that
{\em exactly} simulates all the other well-formed QTMs with
$\appcomplex$-amplitudes.  Instead, we relax the meaning of
``completeness''.

Following Deutsch's work \cite{Deu85}, Bernstein and Vazirani
\cite{BV97} constructed a {\em universal} QTM that can approximately
simulate any well-formed QTM $M$ for $t$ steps with desired accuracy
$\epsilon$ at the cost of polynomial slowdown. In other words, every
QTM can be ``approximately'' reduced to one single QTM. We can
generalize this notion of universality in the following fashion.

\begin{definition}
1) Let $f$ and $g$ be any functions from $\{0,1\}^*$ to $[0,1]$. The
function $f$ is {\em (polynomial-time) approximately reducible to}
$g$, denoted $f\appreduces g$, if there is a function $k \in\fp$ such
that, for every $x$ and $m\in\nat^{+}$, $|f(x) - g(k(x01^m))| \leq
1/m$.

2) Let $\FF$ be any class of functions from $\{0,1\}^*$ to $[0,1]$.  A
function $g$ from $\Sigma^*$ to $[0,1]$ is {\em universal} for $\FF$
(or {\em $\FF$-universal}, in short) if (i) $g$ is in $\FF$ and (ii)
every function $f\in\FF$ is approximately reducible to $g$.
\end{definition}

Unfortunately, we may not replace the term $1/m$ in the above
definition by $2^{-m}$. This relation $\appreduces$ is reflexive and
transitive.  The proof is immediate from the definition.

\begin{lemma}
The relation $\appreduces$ satisfies that (i) for any $f$,
$f\appreduces f$ and (ii) for any $f$, $g$, and $h$, if $f\appreduces
g$ and $g\appreduces h$, then $f\appreduces h$.
\end{lemma}

The importance of a universal function is given as in Lemma
\ref{lemma:universal}. Before describing the lemma, we introduce
useful notations for ``approximate membership'' and ``approximate
inclusion.''

\begin{definition}
Let $\FF$ and $\GG$ be any two classes of functions from $\Sigma^*$ to
the unit real interval $[0,1]$ and let $f$ be any function from
$\Sigma^*$ to $[0,1]$.  The notation $f\appin \FF$ means that, for
every polynomial $p$, there exists a function $g$ in $\FF$ satisfying
$|f(x) - g(x)|\leq 1/p(|x|)$ for all $x$. The notation
$\FF\appsubseteq\GG$ means that $f\appin \GG$ for any function $f$ in
$\FF$.
\end{definition}

\begin{lemma}\label{lemma:universal}
Let $\FF$ and $\GG$ be any two classes of functions from $\Sigma^*$ to
$[0,1]$.  Assume that $\GG\circ\fp\subseteq\GG$ and let $f$ be any
$\FF$-universal function. Then, $\FF\appsubseteq\GG$ iff $f\appin\GG$.
\end{lemma}

\begin{proof}
(Only If - part) This is trivial since $f\in \FF$. (If -part) Assume
that $f\appin\GG$. Take any function $g$ in $\FF$ and any polynomial
$p$. Since $f$ is $\FF$-universal, $g\appreduces f$. There exists a
function $k\in\fp$ such that $|g(x) - f(k(x01^{2p(|x|)}))|\leq
1/2p(|x|)$ for all $x$. Moreover, since $f\appin\GG$, there exists a
function $r\in\GG$ such that $|f(k(x01^{2p(|x|)})) -
r(k(x01^{2p(|x|)}))|\leq 1/2p(|x|)$.  Define
$f'(x)=f(k(x01^{2p(|x|)}))$ and $r'(x)=r(k(x01^{2p(|x|)}))$. Clearly,
$r'$ is in $\GG\circ\fp
\subseteq\GG$. Thus, $g\appin\GG$ since $|g(x) - r'(x)|\leq |g(x)
-f'(x)| + |f'(x) - r'(x)|\leq 1/p(|x|)$. This implies that
$\FF\appsubseteq\GG$.
\end{proof}

Most natural classes satisfy the premise of Lemma
\ref{lemma:universal}. For example, $\sharpqp$, $\gapqp$,
$\optsharpqp$, and $\qoptsharpqp$ satisfy the premise. 

Most well-known quantum complexity classes are believed to lack
complete problems. The notion of universality naturally provides {\em
promise complete} problems for, \eg $\bqp$ by posing appropriate
restrictions on the acceptance probabilities of quantum functions in
$\sharpqp$.

The notion of a universal QTM given by Bernstein and Vazirani
\cite{BV97} gives rise to the $\sharpqp$-universal function QAP. We
assume each DTM has its code (or its description) expressed in
binary. A {\em code} of an amplitude $\alpha$ in $\appcomplex$ is a
code of a DTM that approximates $\alpha$ to within $2^{-n}$ in time
polynomial in $n$. A {\em code} of a QTM means the description of the
machine with codes of all amplitudes used for the QTM. We assume that
any code is expressed in binary.

We fix the universal QTM $M_{U}$ that, on input $\pair{M,x,1^t,1^m}$,
simulates $M$ on input $x$ for $t$ steps and halts in a final
configuration $\qubit{\phi_{M_U}}$ satisfying that
$\|\qubit{\phi_{M_U}} - \qubit{\phi_{M}}\|\leq 1/m$, where
$\qubit{\phi_M}$ is the configuration of $M$ on $x$ after $t$ steps
\cite{BV97}. For completeness, we assume that if $M$ is not a
well-formed QTM then $M_{U}$ rejects the input with probability
$1$. From the construction of a universal QTM in \cite{BV97}, we can
assume that $M_{U}$ has $\{0,\pm1,\pm\cos\theta,\pm\cos\theta,\pm
e^{i\theta}\}$-amplitudes, where
$\theta=2\pi\sum_{i=1}^{\infty}2^{-2^{i}}$. Adleman \etal
\cite{ADH97} further simplified the above set of amplitudes; for
example, we can replace $\cos\theta$ and $\sin\theta$ by $3/5$ and
$4/5$, respectively.

We then define the {\sc Qtm Approximation Problem} (QAP) as
follows:
\ms

\n{\sc Qtm Approximation Problem:} QAP
\vs{-2}
\begin{itemize}
\item {\sf input:} $\pair{M,x,1^t,1^m}$, where $M$ is a  
$\appcomplex$-amplitudes well-formed QTM, $t\in\nat$, and
$m\in\nat^{+}$.
\vs{-2}
\item {\sf output:} the acceptance probability of $M_{U}$ on input 
$\pair{M,x,1^t,1^m}$.
\end{itemize}

As Bernstein and Vazirani \cite{BV97} demonstrated, every function $f$
outputting the acceptance probability of a certain polynomial-time
well-formed QTM with $\appcomplex$-amplitudes is approximately
reducible to QAP.  It is easy to see that QAP is in $\sharpqp$. We
have the following proposition.

\begin{proposition}\label{prop:sharpqp}
QAP is $\sharpqp$-universal.
\end{proposition}

The proof of Proposition \ref{prop:sharpqp} uses the following
folklore lemma. Although the proof of the lemma is easy, it is given
in Appendix for completeness.

\begin{lemma}\label{lemma:epsilon}
Let $M$ and $N$ be two well-formed QTMs. Let $U_{M}$ and $U_{N}$ be
the superpositions of final configurations of $M$ on input
$\qubit{\phi}$ and of $N$ on input $\qubit{\psi}$, respectively.  Let
$\eta_{M}(\qubit{\phi})$ and $\eta_{N}(\qubit{\psi})$ be the
acceptance probabilities of $M$ on input $\qubit{\phi}$ and of $N$ on
input $\qubit{\psi}$, respectively. Then, $|\eta_{M}(\qubit{\phi}) -
\eta_{N}(\qubit{\psi})| \leq \|U_{M}\qubit{\phi} -
U_{N}\qubit{\psi}\|$.
\end{lemma}

Now, we give the proof of Proposition \ref{prop:sharpqp}.

\begin{proofof}{Proposition \ref{prop:sharpqp}}
Let $p$ be any polynomial and let $g$ be any function in
$\sharpqp$. Let $M$ be a polynomial-time $\appcomplex$-amplitude
well-formed QTM that witnesses $g$ with quantum index size $p$. 
Let $q$ be a
polynomial that bounds the running time of $M$. We show that
$g\appreduces \mathrm{QAP}$.

Fix $x$ and $m$ arbitrarily. Let $\qubit{\phi_{M_U}}$ be the final
configuration of $M_{U}$ on input $\pair{M,x,1^{q(|x|)},1^m}$ and
define $\qubit{\phi_{M}}$ to be the final configuration of $M$ on
input $x$. We set $k(x01^m)=\pair{M,x,1^{q(|x|)},1^m}$. It follows
from the definition of $M_{U}$ that $\|\qubit{\phi_{M_{U}}} -
\qubit{\phi_{M}}\|\leq 1/m$. Thus, Lemma
\ref{lemma:epsilon} implies that 
$|\mathrm{QAP}(\pair{M,x,1^t,1^{m}}) - g(x)|\leq
\|\qubit{\phi_{M_U}}-\qubit{\phi_{M}}\|\leq 1/m$. Therefore, we obtain
\[
|\mathrm{QAP}(k(x01^m)) - g(x)| = |\mathrm{QAP}(\pair{M,x,1^t,1^{m}})
- g(x)|\leq \frac{1}{m}.
\]
This guarantees $g\appreduces \mathrm{QAP}$.
\end{proofof}

We exhibit a canonical universal problem for $\qoptsharpqp$. To avoid
any notational inconvenience, we write
$\mathrm{QAP}(\pair{M,x,1^t,1^m},s)$ when $M$ takes an input
$\pair{x,s}$.  Based on this convention, we define the {\sc Maximum
Qtm Problem} (MAXQTM) as follows.
\ms

\n{\sc Maximum Qtm Problem:} MAXQTM
\vs{-2}
\begin{itemize}
\item {\sf input:} $\pair{M,x,1^t,1^m}$, where $M$ is a  
$\appcomplex$-amplitudes well-formed QTM, $t\in\nat$, and
$m\in\nat^{+}$.
\vs{-2}
\item {\sf output:} the maximal acceptance probability, over 
all quantum indices $\qubit{\phi}$ of size $|x|$, of $M_{U}$ on input
$(\pair{M,x,1^t,1^m},\qubit{\phi})$.
\end{itemize}

Obviously, MAXQTM belongs to $\qoptsharpqp$. We further claim that
MAXQTM is indeed universal for $\qoptsharpqp$.

\begin{theorem}
MAXQTM is $\qoptsharpqp$-universal.
\end{theorem}

\begin{proof}
To see this, let $f$ be any function in $\qoptsharpqp$. There exist a
polynomial $p$ and a function $g\in\sharpqp$ such that $f(x) =
\sup\{g(x,\qubit{\phi})\mid \mbox{ $\qubit{\phi}$ is any qustring of
size $p(|x|)$ }\}$ for every $x$. Without loss of generality, assume
that $p(n)>n+1$ for all $n$. Take an appropriate QTM $M_g$ that
witnesses $g$.  We show that $f\appreduces
\mathrm{MAXQTM}$. Define another QTM $M_g'$ as follows: on
input $z$ and $s\in\Sigma^{|z|}$, if $z$ is not of the form
$x01^{p(|x|)-|x|-1}$, then reject the input. Otherwise, simulate $M_g$
on input $(x,s)$.

Let $q$ be any polynomial such that, for every $x$, $M_g'$ on input
$x01^{p(|x|)-|x|-1}$ halts in at most $q(|x|)$ steps. Thus,
Proposition \ref{prop:sharpqp} implies that, for every
$s\in\Sigma^{p(|x|)}$,
\[
 |\mathrm{QAP}(\pair{M_g',x01^{p(|x|)-|x|-1},1^{q(|x|)},1^m},s)
 - g(x,s)| \leq \frac{1}{m}.
\]
This clearly yields the following inequality:
\[
 |\sup_{\qubit{\phi}}
\{\mathrm{QAP}(\pair{M_g',
x01^{p(|x|)-|x|-1},1^{q(|x|)},1^m},\qubit{\phi})\}
- \sup_{\qubit{\phi}}\{g(x,\qubit{\phi})\}|\leq \frac{1}{m},
\]
where $\qubit{\phi}$ runs over all qustrings of size $p(|x|)$.
To complete the proof, it suffices to define $k(x01^m) =
\pair{M_g',x01^{p(|x|)-|x|-1},1^{q(|x|)},1^m}$ for
all $x$ and $m$.
\end{proof}

\begin{corollary}
$\mathrm{MAXQTM}\appin \sharpqp$ iff $\qoptsharpqp\appsubseteq\sharpqp$.
\end{corollary}

We have just shown an example of universal optimization problem,
MAXQTM, for $\qoptsharpqp$. Nonetheless, MAXQTM heavily relies on a
universal QTM and it seems artificial. To develop a fruitful theory of
quantum optimization problems, we need ``natural'' examples of
universal problems in a variety of fields, such as graph theory,
logic, and group theory.  Finding such natural problems is one of the
most pressing open problems in our theory.

\section{Relationship between Qopt\#QP and \#QP}
\label{sec:function-class}

The quantum optimization problems $\qoptsharpqp$ are induced from
$\sharpqp$-functions by taking quantum indices. However, there has
been shown few relationship between $\qoptsharpqp$ and $\sharpqp$
except for the trivial inclusion $\sharpqp\subseteq\qoptsharpqp$. To
fill in the gap between them, we begin with a simple observation using
Lemma \ref{lemma:property-1}(4) (see also \cite{Wat01}).
 
\begin{proposition}\label{prop:approximation}
For any function $h$ in $\qoptsharpqp$, there exists a function
$g\in\sharpqp$ and a polynomial $p$ such that, for all strings $x$ and
integers $m>0$,
\[
  g(x01^m)\leq h^{m}(x) \leq 2^{p(|x|)}g(x01^m).
\]
\end{proposition} 

\begin{proof}
Let $h$ be any function in $\qoptsharpqp$. Let
$h'(\pair{x,y})=h(x)^{|y|}$. By Lemma \ref{lemma:property-1}(4), $h'$
belongs to $\qoptsharpqp$. The key idea comes from the fact that, in
the proof of Lemma \ref{lemma:property-1}(4), we can assume that, for
a certain fixed polynomial $p$ and a function $f\in\sharpqp$,
$h'(\pair{x,y})=\sup\{f(\pair{x,y},\qubit{\phi})\mid
\mbox{ $\qubit{\phi}$ is a qustring of size ${p(|x|)}$ }\}$ for all 
$x$ and $y$. Notice that the size of $\qubit{\phi}$ is independent of
$y$.

The rest of the proof is to estimate $h'(\pair{x,y})$.  Define
$g(x01^m)=2^{-p(|x|)}\sum_{s:|s|=p(|x|)}f(\pair{x,1^m},s)$ for all $x$
and $m$. Obviously, $g\in\sharpqp$. By the maximality of $h'$,
$f(\pair{x,1^m},s)\leq h'(\pair{x,1^m})$ for every
$s\in\Sigma^{p(|x|)}$. Since the value $2^{-p(|x|)}
\sum_{s:|s|=p(|x|)}f(\pair{x,1^m},s)$ is the average of all
$f(\pair{x,1^m},s)$, we have $2^{-p(|x|)}
\sum_{s:|s|=p(|x|)}f(\pair{x,1^m},s) \leq h'(\pair{x,1^m})$, 
which implies $g(x01^{m})\leq
h'(\pair{x,1^m})$. On the other hand, $h'(\pair{x,1^m})\leq
\sum_{s:|s|=p(|x|)}f(\pair{x,1^m},s)$, which implies $h'(\pair{x,1^m})\leq
2^{p(|x|)}g(x01^m)$.
\end{proof}

Although Proposition \ref{prop:approximation} is a rough
approximation, it is used as a tool in Section \ref{sec:definable}.

We further explore a relationship between $\qoptsharpqp$ and
$\sharpqp$. In the previous section, we have introduced the notations
$\appin$ and $\appsubseteq$ in connection to universal problems. These
notations are used to indicate a certain notion of ``closeness'' (that
is, bounded by reciprocal of a polynomial). In this section, we need a
much tighter notion of ``closeness.''

\begin{definition}
Let $\FF$ and $\GG$ be any two classes of functions. For any function
$f$, we write $f\stappin\FF$ if, for every polynomial $p$, there
exists a function $g\in\FF$ such that, for every $x$, $|f(x)-g(x)|\leq
2^{-p(|x|)}$. The notation $\FF\stappsubseteq\GG$ means that
$f\stappin \GG$ for all functions $f$ in $\FF$.
\end{definition}

In Proposition \ref{prop:inclusion}, we show that any quantum
optimization problem in $\qoptsharpqp$ can be closely approximated by
$\sharpqp$-functions with the help of certain oracles. For this
purpose, we define the new class, called $\qop$. The following
definition resembles the $\sharpqp$-characterization of $\pp$ sets in
Lemma \ref{lemma:pp-sharpqp}.

\begin{definition}\label{def:qop}
A set $A$ is in $\qop$ (``quantum optimization polynomial time'') if
there exist two functions $f,g\in\qoptsharpqp$ and a function
$h\in\fp$ such that, for every $x$, $x\in A$ exactly when
$\floors{2^{|h(x)|}f(x)}> \floors{2^{|h(x)|}g(x)}$. This $h$ is called
a {\em selection function}.
\end{definition}

Note that, in Definition \ref{def:qop}, we can replace the value
$|h(x)|$ by $p(|x|)$ for an appropriate polynomial $p$. This is seen
by taking $p$ that satisfies $|h(x)|\leq p(|x|)$ for all $x$ and by
replacing $f$ and $g$, respectively, with
$\hat{f}(x)=2^{-p(|x|)+|h(x)|}f(x)$ and
$\hat{g}(x)=2^{-p(|x|)+|h(x)|}g(x)$.  It also follows from Proposition
\ref{prop:AQMA-WQPP} that $\pp\subseteq\qop$.

In the following proposition, we view an $\fpspace$-function as a
function from $\Sigma^*$ to $\dyadic$.

\begin{proposition}\label{prop:inclusion}
$\qoptsharpqp\stappsubseteq\sharpqp^{\qop}
\stappsubseteq\fpspace\cap\dyadic^{\Sigma^*}$.
\end{proposition}

\begin{proof}
Let $f\in\qoptsharpqp$. Take a function $g\in\sharpqp$ and a
polynomial $q$ such that $f(x)=\sup\{g(x,\qubit{\phi})\mid \mbox{
$\qubit{\phi}$ is any qustring of size $q(|x|)$ }\}$ for every
$x$. Define the sets $A_{+}$ and $A_{-}$ as follows:
$A_{+}=\{(x,y)\mid
\floors{2^{p(|x|)+1}f(x)}\geq num(y)\}$ and $A_{-}=\{(x,y)\mid 
\floors{2^{p(|x|)+1}f(x)}\leq num(y)\}$, where $num(y)$ is the 
number $n$ in $\nat$ such that $y$ is the $n$th string in the standard
order (the empty string is the $0$th string). Let $A=A_{+}\oplus
A_{-}$, where $\oplus$ is the disjoint union. It is not difficult to
show that $A\in\qop$. Let $p$ be any polynomial. By a standard binary
search algorithm using $A$, we can find an approximation of $f(x)$ to
within $2^{-p(|x|)}$ in polynomial time. Let $h^A$ be the function
computed by this algorithm with oracle $A$. Then, we have
$|f(x)-h^A(x)|\leq 2^{-p(|x|)}$.

For the second inclusion, it suffices to show that
$\qoptsharpqp\stappsubseteq\fpspace\cap\dyadic^{\Sigma^*}$ since this
implies $\qop\subseteq\pspace$ together with the fact that
$\sharpqp\stappsubseteq \fpspace\cap\dyadic^{\Sigma^*}$ (which follows
from the result in \cite{Yam99b}).  Let $f$ be any function in
$\qoptsharpqp$.  Let $M$ be a QTM that witnesses $f$ with a polynomial
$p$ bounding the quantum index size. Recall the QTM $N_{M,p}$ and its
associated matrices $\{P_{M,p,x}\}_{x\in\Sigma^*}$ given in Section
\ref{sec:quantum-opt} such that, for every $x$, (i) $P_{M,p,x}$ is a
$2^{p(|x|)}\times2^{p(|x|)}$ positive semidefinite, contractive,
Hermitian matrix and (ii) $f(x)=\|P_{M,p,x}\|$.  Note that each
$(s,t)$-entry of $P_{M,p,x}$ is approximated to within $2^{-m}$
deterministically using space polynomial in $m$, $|x|$, $|s|$, and
$|t|$ by simulating all computation paths of $N_{M,p}$ on input
$(x,t)$. Consider the characteristic polynomial $u(z)= \mathrm{det}(zI
- V_x)$.  This $u(z)$ has the form
$u(z)=\sum_{i=0}^{2^{p(|x|)}}a_{i}z^i$. It is easy to see that each
coefficient $a_{i}$ can be approximated using polynomial space (see
also the argument in \cite{Wat02}). Thus, we can also approximate each
real root of $u(z)$ using polynomial space. Finally, we find a good
approximation of the maximal eigenvalue of $P_{M,p,x}$. Thus, $f\appin
\fpspace\cap\dyadic^{\Sigma^*}$.
\end{proof}

The following is an immediate consequence of Proposition
\ref{prop:inclusion}.

\begin{corollary}\label{cor:collapse}
If $\eqp=\qop$, then $\qoptsharpqp\stappsubseteq\sharpqp$.
\end{corollary}

The converse of Corollary \ref{cor:collapse} does not seem to
hold. Instead, we show the following weaker form of the converse. For
our proposition, let $\widehat{\qop}$ denote the subset of $\qop$ with
the extra condition that $\floors{2^{|h(x)|}f(x)}\neq
\floors{2^{|h(x)|}g(x)}$ for all $x$ in Definition \ref{def:qop}.

\begin{proposition}
If $\qoptsharpqp\stappsubseteq\sharpqp$, then $\widehat{\qop}=\pp$ and
$\qop\subseteq\p^{\sharpp[1]}$, where $[\cdot]$ stands for the number
of queries on each input.
\end{proposition}

\begin{proof}
Assume that $\qoptsharpqp\stappsubseteq\sharpqp$. We first show that
$\widehat{\qop}=\pp$. Let $A$ be any set in $\widehat{\qop}$. There
exist a polynomial $p$ and functions $f$ and $g$ in $\optsharpqp$
such that, for every $x$, (i) if $x\in A$ then
$\floors{2^{p(|x|)}f(x)} > \floors{2^{p(|x|)}g(x)}$ and (ii) if
$x\not\in A$ then $\floors{2^{p(|x|)}f(x)} <
\floors{2^{p(|x|)}g(x)}$. Since $f\appin\sharpqp$ by our assumption,
there exists a certain function $\hat{f}$ in $\sharpqp$ such that, for
every $x$, $|f(x)-\hat{f}(x)|\leq 2^{-p(|x|)-1}$, which implies that
$\floors{2^{p(|x|)}f(x)}=\floors{2^{p(|x|)}\hat{f}(x)}$. Similarly, we
obtain $\hat{g}$ from $g$. Assume that $x\in A$. Then, we have
$\floors{2^{p(|x|)}f(x)}>\floors{2^{p(|x|)}g(x)}$. This clearly
implies $\hat{f}(x)>\hat{g}(x)$. Similarly, if $x\not\in A$ then
$\hat{f}(x)<\hat{g}(x)$. Therefore, $A=\{x\mid
\hat{f}(x)>\hat{g}(x)\}$. By Lemma \ref{lemma:pp-sharpqp}, $A$ belongs
to $\pp$.
 
Next, we show that $\qop\subseteq\p^{\sharpp[1]}$. Let $A$ be in
$\qop$. Similar to 1), we obtain $\hat{f}$ and $\hat{g}$ in
$\sharpqp$. For this $\hat{f}$, by \cite{Yam99b}, there exist two
functions $\tilde{f}\in\gapp$ and $\ell\in\fp$ such that, for every
$x$, $|\hat{f}(x) - \tilde{f}(x)/\ell(1^{|x|})|\leq
2^{-p(|x|)-1}$. Thus, $\floors{2^{p(|x|)}\hat{f}(x)}$ coincides with
the first $p(|x|)$ bits of the binary expansion of
$2^{p(|x|)}\tilde{f}(x)/\ell(1^{|x|})$. Similarly, we obtain
$\tilde{g}$ and $\ell'$. Without loss of generality, we can assume
that $\ell=\ell'$.  To determine whether $x$ is in $A$, we first
compute the value $\tilde{f}(x)$ and $\tilde{g}(x)$ and then compare
the first $p(|x|)$ bits of the binary expansion of
$2^{p(|x|)}\tilde{f}(x)/\ell(1^{|x|})$ and those of
$2^{p(|x|)}\tilde{g}(x)/\ell(1^{|x|})$. This is done by making one
query to each of $\tilde{f}$ and $\tilde{g}$. It is easy to reduce
these two queries to one single query; for example, we can define
another $\sharpp$-function $h(x)$ to be
$2^{q(|x|)}\tilde{f}(x)+\tilde{g}(x)$, where $q$ is a polynomial
satisfying $|\tilde{f}(x)|+|\tilde{g}(x)|\leq q(|x|)$ for all
$x$. Thus, $A$ belongs to $\p^{\sharpp[1]}$.
\end{proof}

By a simple argument, we next show that $\bqp\neq\qma$ implies
$\qoptsharpqp\not\appsubseteq\sharpqp$. We use the result from Lemma
\ref{lemma:definable}(2) that $A\in\qma$ iff there exists a
$f\in\qoptsharpqp$ such that, for every $x$, if $x\in A$ then
$f(x)\geq 3/4$ and otherwise $f(x)\leq 1/4$.

\begin{lemma}\label{lemma:opt-vs-sharpqp}
$\qoptsharpqp\not\appsubseteq\sharpqp$ unless $\bqp=\qma$.
\end{lemma}

\begin{proof}
Assume that $\qoptsharpqp\appsubseteq\sharpqp$. Let $A$ be any set in
$\qma$. Using the amplification property of $\qma$ (see \cite{Wat00}),
we can assume the existence of a function $f$ in $\qoptsharpqp$ such
that, for every $x$, if $x\in A$ then $f(x)\geq 7/8$ and otherwise
$f(x)\leq 1/8$.  By our assumption, there exists a function
$g\in\sharpqp$ such that $|f(x)-g(x)|\leq 1/8$. Thus, if $x\in A$ then
$g(x)\geq f(x)-\frac{1}{8}\geq
\frac{3}{4}$. If $x\not\in A$ then $g(x)\leq
f(x)+\frac{1}{8} \leq \frac{1}{4}$.  Therefore, $A$ belongs to $\bqp$.
\end{proof}

Recall the fact that $\np^A\nsubseteq\bqp^A$ for a certain set $A$
\cite{BV97,BBBV97} and that $\np^B\subseteq\qma^B$ for any set $B$. As
a consequence, we obtain the following corollary since Lemma
\ref{lemma:opt-vs-sharpqp} relativizes.

\begin{corollary}
There exists a set $B$ such that
$\qoptsharpqp^B\not\appsubseteq\sharpqp^B$.
\end{corollary}

\section{Qopt\#QP-Definable Classes}\label{sec:definable}

This section discusses the relationship between quantum optimization
problems and known complexity classes. To describe the relationship,
one useful notion is ``definability'' of Fenner \etal \cite{FFK94},
which is originally defined only for $\gapp$-functions. The
$\sharpqp$-definability is also discussed in \cite{Yam99b}. We give
this ``definability'' below in a general fashion.

\begin{definition}
Let $\FF$ be any set of functions from $\Sigma^*$ to $\real$. A class
$\CC$ of sets is said to be {\em (uniformly) $\FF$-definable} if there
exist an positive integer $k$ and a pair of disjoint sets
$A,R\subseteq\Sigma^*\times\real$ such that, for every set $A$ in
$\CC$, there exists a function $f$ in $\FF$ satisfying the following
condition: for every $x$, if $x\in A$ then $(x,f(x))\in A$ and
otherwise $(x,f(x))\in R$.
\end{definition}

In what follows, we focus on $\qoptsharpqp$-definable classes. Obvious
examples of such classes are $\eqma$ and $\qma$, where $\eqma$ is an
error-free restriction of $\qma$ \cite{KMY01} defined as follows.  A
set $S$ is in $\eqma$ if there exists a polynomial $p$ and a
polynomial-time $\appcomplex$-amplitude well-formed QTM $M$ such that,
for every $x$, (i) if $x\in S$ then there exists a qustring
$\qubit{\phi_x}$ of size $p(|x|)$ that forces $M$ to accept
$\qubit{x}\qubit{\phi_x}$ with certainty and (ii) if $x\not\in S$
then, for any qustring $\qubit{\phi}$ of size $p(|x|)$, $M$ rejects
$\qubit{x}\qubit{\phi}$ with certainty.

\begin{lemma}\label{lemma:definable}
1. $\eqma$ is $\qoptsharpqp$-definable. More precisely, $A\in\eqma$
iff $\chi_{A}\in\qoptsharpqp$, where $\chi_{A}$ is the {\em
characteristic function} of $A$; that is, $\chi_{A}(x)=1$ if $x\in A$
and $\chi_{A}(x)=0$ otherwise.

2. $\qma$ is $\qoptsharpqp$-definable. More strongly, a set $S$ is in
$\qma$ iff there exist a polynomial $p$ and a function
$f\in\qoptsharpqp$ such that, for every $x$, if $x\in A$ then
$f(x)\geq 3/4$ and otherwise $f(x)\leq 1/4$.
\end{lemma}

A less obvious example of $\qoptsharpqp$-definable class
is $\nqp$. This is an
immediate consequence of Proposition \ref{prop:approximation}.

\begin{lemma}
$\nqp$ is $\qoptsharpqp$-definable. More strongly, a set $A$ is in
$\nqp$ iff there exists a function $f\in\qoptsharpqp$ such that, for
every $x$, (i) if $x\in A$ then $f(x)>0$, and (ii) if $x\not\in A$
then $f(x)=0$.
\end{lemma}

\begin{proof}
(If -- part) Assume that $A$ and $f$ in $\qoptsharpqp$ satisfy that,
for every $x$, $x\in A$ iff $f(x)>0$. By Proposition
\ref{prop:approximation}, there exists a function $g\in\sharpqp$ such
that, for every $x$, $f(x)>0$ iff $g(x)>0$. Thus, $A\in\nqp$.  (Only
if -- part) This is trivial because $\sharpqp\subseteq\qoptsharpqp$.
\end{proof}

The class $\pp$ is one of the most important complexity classes and
$\pp$ also contains most well-known quantum complexity classes, such as
$\bqp$, $\qma$, and $\nqp$. Moreover, $\pp$ is robust in the sense
that it equals $\pqp_{\appcomplex}$, a quantum interpretation of $\pp$
\cite{Yam99b}. Thus, it is natural to ask whether $\pp$ is
$\qoptsharpqp$-definable.  The following proposition is a partial
answer to this question. This proposition also yields Watrous's recent
result that $\qma\subseteq\pp$ \cite{Wat01}.

\begin{proposition}\label{prop:PP}
Let $A$ be any subset of $\{0,1\}^*$. The following statements are
equivalent.
\vs{-2}
\begin{enumerate}
\item $A$ is in $\pp$.
\vs{-2}
\item For every polynomial $q$, there exist two functions 
$f\in\qoptsharpqp$ and $g\in\gapqp$ such that, for every string $x$
and integer $m$ ($m\geq|x|$), (i) $g(x01^m)>0$; (ii) $x\in A$ implies
$(1-2^{-q(m)})g(x01^m)\leq f(x01^m)\leq g(x01^m)$; and (iii) 
$x\not\in A$ implies $0\leq f(x01^m)\leq 2^{-q(m)}g(x01^m)$.
\end{enumerate}
\end{proposition}

\begin{proof}
(1 implies 2) Let $A$ be any set in $\pp$. Note that 
the following characterization of $\pp$ is 
well-known (see \cite{Li93,HO02}).

\begin{fact}\label{fact:PP}\hs{2}
A set is in $\pp$ iff, for every polynomial $p$, there exist two
functions $f,g\in\gapp$ such that, for every $x$ and every $m\geq|x|$,
(i) $g(x01^m)>0$, (ii) $x\in A$ implies $(1-2^{-q(m)})g(x01^m)\leq
f(x01^m)\leq g(x01^m)$, and (iii) $x\not\in A$ implies $0\leq
f(x01^m)\leq 2^{-q(m)}g(x01^m)$.
\end{fact}
Take $\tilde{g}$ and $\tilde{f}$ from $\gapqp$ and $\ell$ from $\fp$
such that $f(x)=\tilde{f}(x)\ell(1^{|x|})$ and
$g(x)=\tilde{g}(x)\ell(1^{|x|})$ for all $x$ \cite{Yam99b}. We thus
replace $g$ and $f$ in Fact \ref{fact:PP} by $\tilde{g}$ and
$\tilde{f}$ and then make all the terms squared. Note that
$\tilde{f}^2$ and $\tilde{g}^2$ belong to $\sharpqp$ \cite{Yam99b} and
thus, the both are in $\qoptsharpqp$.  Since $(1-2^{-q(m)})^2\geq
1-2^{-q(m)+1}$ and $(2^{-q(m)})^2\leq 2^{-q(m)+1}$, we can obtain the
desired result.

(2 implies 1) Set $q(n)=n$ and assume that there exist two functions
$f\in\qoptsharpqp$ and $g\in\gapqp$ satisfying the conditions of
statement 2). Since $g\in\gapqp$ implies $g^2\in\sharpqp$
\cite{Yam99b}, we can assume from the beginning of this proof that
$g\in\sharpqp$. For each $x$, let $\hat{x}=x01^{|x|}$.  Proposition
\ref{prop:approximation} guarantees the existence of a polynomial $p$
and a function $h\in\sharpqp$ satisfying that, for every $x$ and
$m>1$, $h(\hat{x}01^m)\leq f^m(\hat{x})\leq 2^{p(|x|)}h(\hat{x}01^m)$.
We can also assume that $p(n)\geq2$ for all $n\in\nat$.  Let
$g'(x)=2^{-2p(|x|)}g^{p(|x|)}(\hat{x})$ and
$h'(x)=h(\hat{x}01^{p(|x|)})$ for all $x$. It is easy to see that
$g',h'\in\sharpqp$ since $g,h\in\sharpqp$. In what follows, we show
that, for all but finitely-many strings $x$, $x\in A$ iff $g'(x) <
h'(x)$. This implies that, by Lemma \ref{lemma:pp-sharpqp}, $A$ is
indeed in $\pp$.

Fix $x$ arbitrarily with $n=|x|\geq4$. In the case where $x\in A$, we
obtain:
\[
2^{-\frac{p(n)}{n-2}}g^{p(n)}(\hat{x}) \leq
(1-2^{-n})^{p(n)}g^{p(n)}(\hat{x})
\leq f^{p(n)}(\hat{x}) \leq 2^{p(n)}h(\hat{x}01^{p(n)})
\]
since $(1-2^{-n})^{p(n)} = ((1-2^{-n})^{n-2})^{\frac{p(n)}{n-2}}$ and
$(1-2^{-n})^{n-2}\geq 1-2^{-n+(n-2)+1}=\frac{1}{2}$. This yields the
inequality $2^{-p(n)(1+1/(n-2))}g^{p(n)}(\hat{x}) \leq h'(x)$, which
further implies that $g'(x) < h'(x)$ since $n\geq4$.  Consider the
other case where $x\not\in A$. In this case, we obtain:
\[
h'(x)= h(\hat{x}01^{p(n)})\leq f^{p(n)}(\hat{x})\leq 
(2^{-n})^{p(n)}g^{p(n)}(\hat{x})
\leq 2^{-np(n)}g^{p(n)}(\hat{x}),
\]
which implies $h'(x) < 2^{-2p(n)}g^{p(n)}(\hat{x}) = g'(x)$ since
$n>2$.
\end{proof}

\section{Complexity Classes Induced from Qopt\#QP}
\label{sec:QOP}

The $\qoptsharpqp$-definability is a tool in demonstrating a close
relationship between $\qoptsharpqp$ and other complexity
classes. Here, we take another approach toward the study of the
complexity of quantum optimization problems. Instead of
$\qoptsharpqp$, we explore complexity classes induced from
$\qoptsharpqp$. We have already seen the class $\qop$, which has
played an important role in approximating quantum optimization
problems in Proposition \ref{prop:inclusion}.

The following claim for $\qop$ is trivial
since $\qoptsharpqp\circ \fp = \qoptsharpqp$.

\begin{lemma}
$\qop$ is closed downward under polynomial-time many-one reductions.
\end{lemma}

We next focus on sets with low information, known as low sets, for
$\qop$. Here is a general definition of $\FF$-low sets.

\begin{definition}
Let $\FF$ be any class (of functions or sets). For any set $A$, $A$ is
{\em low} for $\FF$ (or {\em $\FF$-low}) if $\FF^A=\FF$. Let $\low\FF$
denote the collection of all sets that are $\FF$-low.
\end{definition}

An example of such a low set is $\low\sharpqp=\low\gapqp=\eqp$
\cite{Yam99b}.  In the following proposition, we show the low sets for
$\qoptsharpqp$ and $\qop$.

\begin{proposition}
1. $\eqp\subseteq \low\qoptsharpqp\subseteq \eqma$.

2. $\low\qoptsharpqp \subseteq \low\qop \subseteq \qop\cap\co\qop$.
\end{proposition}

\begin{proof}
1) For the first inclusion, assume that $A\in\eqp$. Let $M_{A}$ be its
associated QTM. Using the squaring lemma in \cite{Yam99a}, we can
assume that $M_{A}$ reaches a specific final configuration (in which
the tape content is either $\qubit{x}\qubit{1}$ or
$\qubit{x}\qubit{0}$ and empty elsewhere) with probability
$1$. Consider any function $f$ in $\qoptsharpqp^A$ and its associated
QTM $M$. As shown in \cite{Yam99a}, we can also assume that we have
the same number of queries on each computation path of $M$. We then
obtain a non-relativized QTM by substitute the simulation of $M_{A}$
for a query given by $M$. Since this new machine witnesses $f$, $f$
belongs to $\qoptsharpqp$.

For the second inclusion, assume that
$\qoptsharpqp^A=\qoptsharpqp$. Since $\chi_{A}\in\qoptsharpqp^A$, we
conclude that $\chi_{A}\in\qoptsharpqp$, which implies that
$A\in\eqma$ by Lemma \ref{lemma:definable}(1).

2) The first inclusion is trivial. The second inclusion comes from the
fact that $A\in \qop^A\cap\co\qop^A$ for all $A$'s.
\end{proof}

The difference between $\low\qoptsharpqp$ and $\low\qop$ stems from
the floor-operation (rounding down the terms $2^{|h(x)|}f(x)$ and
$2^{|h(x)|}g(x)$) for $\qop$ in Definition \ref{def:qop}.

Another example of $\qoptsharpqp$-related class comes from Proposition
\ref{prop:PP}. Our complexity class is inspired by Li's earlier
work. Li \cite{Li93} introduced the complexity class $\app$: a set $A$
is in $\app$ (``amplified'' $\pp$) iff, for every polynomial $p$,
there exist two functions $f,g\in\gapp$ such that, for every $x$ and
every $m\geq|x|$, (i) $g(1^m)>0$; (ii) if $x\in A$ then
$(1-2^{-q(m)})g(1^m)\leq f(x01^m)\leq g(1^m)$; and (iii) if $x\not\in
A$ then $0\leq f(x01^m)\leq 2^{-q(m)}g(1^m)$.  Now, we introduce its
quantum extension using $\qoptsharpqp$.

\begin{definition}\label{def:aqma}
A set $A$ is in $\aqma$ (``amplified'' QMA) if, for every polynomial $q$,
there exist two functions $f\in\qoptsharpqp$ and $g\in\gapqp$ such
that, for every string $x$ and integer $m$ ($m\geq|x|$),

i) $g(1^m)>0$; 

ii) if $x\in A$ then $(1-2^{-q(m)})g(1^m)\leq f(x01^m)\leq g(1^m)$;
and

iii) if $x\not\in A$ then $0\leq f(x01^m)\leq 2^{-q(m)}g(1^m)$.
\end{definition}

As shown in the proof of Proposition \ref{prop:PP}, by squaring
formulas (i) and (ii) in Definition \ref{def:aqma} (as well as
choosing $q(m)+1$), we can replace the class $\gapqp$ by $\sharpqp$.

Similar to $\app$, the class $\aqma$ also enjoys simple closure
properties: intersection and disjoint union.

\begin{lemma}
$\aqma$ is closed under intersection and disjoint union.
\end{lemma}

\begin{proof}
Let $A_1$ and $A_2$ be any two sets in $\aqma$. Also let $q$ be any
polynomial. For each $i\in\{1,2\}$, there exist two functions
$f_i\in\qoptsharpqp$ and $g_i\in\gapqp$ such that, for every $x$ and
$m\geq|x|$,
\begin{itemize}
\item[(i)] if $x\in A_i$ then 
$(1-2^{-q(m)})g_i(1^m)\leq f_i(x01^m)\leq g_i(1^m)$ and 
\vs{-2}
\item[(ii)] if $x\not\in A_i$ then 
$0\leq f_i(x01^m)\leq 2^{-q(m)}g_i(1^m)$. 
\end{itemize}
By the remark after Definition \ref{def:aqma}, we can assume that
$g_1,g_2\in\sharpqp$.

For the intersection, note that (i') if $x\in A_1\cap A_2$ then
$(1-2^{-q(m)})^2g_1(1^m)g_2(1^m)\leq f_1(x01^m)f_2(x01^m)\leq
g_1(1^m)g_2(1^m)$ and (ii') if $x\not\in A_1\cap A_2$ then $0\leq
f_1(x01^m)f_2(x01^m)\leq (2^{-q(m)})^2 g_1(1^m)g_2(1^m)$. We define
$g(z)=g_1(z)g_2(z)$ and $f(z)=f_1(z)f_2(z)$. Obviously, $g\in\gapqp$
and $f\in\qoptsharpqp$. Moreover, $(1-2^{-q(m)})^2\geq 1-2^{-q(m)+1}$
and $(2^{-q(m)})^2\leq 2^{-q(m)+1}$. Thus, it suffices to take
$q(m)+1$ for $q(m)$ in (i) and (ii).

For the disjoint union, we first multiply the formulas in both (i) and
(ii) by $g_{\bar{i}}(1^m)$, where $\bar{i}=3-i$. Define
$g(z)=g_1(z)g_2(z)$ for all $z$. Obviously, $g$ is in
$\sharpqp$. Also, define $f'$ and $f''$ so that
$f'(x01^m)=f_1(x01^m)g_2(1^m)$ and
$f''(x01^m)=g_1(1^m)f_2(x01^m)$. Clearly, $f',f''\in\qoptsharpqp$
since $g_1,g_2\in\sharpqp\subseteq\qoptsharpqp$. Let $M$ and $M'$ be
appropriate QTMs that witness $f'$ and $f''$, respectively, with a
polynomial $p$ bounding the size of quantum indices. Consider the
following QTM $M$: on input $(bx, \qubit{\phi})$, where $b\in\{0,1\}$,
$x\in\Sigma^*$, and $\qubit{\phi}$ is any quantum index of size
$p(|x|+1)$, if $b=0$ then simulate $M'$ on input $(x,\qubit{\phi})$;
otherwise, simulate $M''$ on input $(x,\qubit{\phi})$. Let $f(z)$ be
the maximal acceptance probability, over all $\qubit{\phi}$, of $M$ on
input $(x,\qubit{\phi})$. It is easy to show that $g$ and $f$ witness
$A_1\oplus A_2$.
\end{proof}

The relationships among aforementioned complexity classes are
summarized as follows. Watrous's recent result $\qma\subseteq\pp$
\cite{Wat01} is also improved to $\qma\subseteq\aqma$.

\begin{proposition}\label{prop:AQMA-WQPP}
$\app\cup\qma\subseteq \aqma \subseteq \pp\subseteq \qop \subseteq
\pspace$.
\end{proposition}

\begin{proof}
It is easy to show that $\app\subseteq\aqma$.  We next show that
$\qma\subseteq\aqma$. Let $A$ be any set in $\qma$. Note that $\qma$
enjoys the amplification property. Let $q$ be any (nondecreasing)
polynomial. There exists a function $f\in\qoptsharpqp$ such that, for
every $x$, if $x\in A$ then $f(x)\geq 1-2^{-q(|x|)}$ and otherwise
$f(x)\leq 2^{-q(|x|)}$.  Define $g(x)=1$ for all strings $x$. It
follows that $f$ and $g$ satisfy Definition \ref{def:aqma}. The
inclusion $\aqma\subseteq\pp$ easily follows from Proposition
\ref{prop:PP}.  Finally, we show that $\pp\subseteq\qop$. This is
obvious by Lemma \ref{lemma:pp-sharpqp} and the following fact: for
every $f\in\sharpqp$, there exists a polynomial $p$ such that
$\floors{2^{p(|x|)}f(x)}=2^{p(|x|)}f(x)$ for all $x$. The last
inclusion $\qop\subseteq\pspace$ follows from Proposition
\ref{prop:inclusion}.
\end{proof}

\bs
\subsection*{Acknowledgments}
The author is truly grateful of Hirotada Kobayashi for his insightful
comments and corrections of errors in an early draft of this paper and
of Steve Homer and Frederic Green for an interesting discussion at
Boston University on quantum optimization problems. He also thanks
Jack C.H. Lin for pointing out typos.

\bibliographystyle{alpha}


\section*{Appendix}

We show from \cite{Yam99a} three local
requirements on a quantum transition function 
that its associated QTM
should be well-formed. 

We begin with the simple case of 1-tape QTMs. Let
$M=(Q,\{q_0\},Q_f,\Sigma,\delta)$ be a 1-tape QTM with initial state
$q_0$ and a set $Q_f$ of final states.  As noted before, the head move
directions $R$, $N$, and $L$ are freely identified with $-1$, $0$, and
$+1$, respectively. We introduce new notation that is helpful to
describe the well-formedness of QTMs.  Let $D=\{0,1,-1\}$,
$E=\{0,\pm1,\pm2\}$, and $H=\{0,\pm1,\natural\}$. For $d\in D$ and
$\epsilon\in E$, we write $E_{d}=\{\epsilon\in E\mid |2d-\epsilon|\leq
1\}$ and $D_{\epsilon}=\{d\in D\mid |2d-\epsilon|\leq 1\}$. Let
$(p,\sigma,\tau)\in Q\times \Sigma^2$ and $\epsilon\in E$.

Bernstein and Vazirani \cite{BV97} introduced the notation
$\delta(p,\sigma|\tau,d)$ for $\sum_{q\in
Q}\delta(p,\sigma,q,\tau,d)\qubit{q}$.  Instead of their notation, we
introduce the notation $\delta[p,\sigma,\tau|\epsilon]$. An element
$\delta[p,\sigma,\tau|\epsilon]$ in $\complex^{H}$ is defined as:
\[
 \delta[p,\sigma,\tau|\epsilon] = \sum_{q\in Q}\sum_{d\in D_{\epsilon}}
 \frac{\delta(p,\sigma,q,\tau,d)}{\sqrt{|E_d|}}
 \qubit{q}\qubit{h_{d,\epsilon}},
\]
where $h_{d,\epsilon}=2d-\epsilon$ if $\epsilon\neq 0$ and
$h_{d,\epsilon}=\natural$ otherwise. The term $\sqrt{|E_d|}$ in the
above equation ensures that $\sum_{\epsilon\in
E}\|\delta[p,\sigma,\tau|\epsilon]\|^2 = \sum_{d\in
D}\|\delta(p,\sigma|\tau,d)\|^2$. 

\begin{lemma}\label{lemma:well-formedness}
A 1-tape QTM $M=(Q,\{q_0\},Q_f,\Sigma,\delta)$ is well-formed iff the
following three conditions hold.
\begin{enumerate}
\item (unit length) $\|\delta(p,\sigma)\|=1$ for all $(p,\sigma)\in
Q\times\Sigma$.

\item (orthogonality) $\delta(p_1,\sigma_1)
\cdot\delta(p_2,\sigma_2)=0$ for any distinct pair $(p_1,\sigma_1),
(p_2,\sigma_2)\in Q\times\Sigma$.

\item (separability) $\delta[p_1,\sigma_1,\tau_1|\epsilon]\cdot
\delta[p_2,\sigma_2,\tau_2|\epsilon']=0$ for any distinct pair
$\epsilon,\epsilon'\in E$ and for any pair
$(p_1,\sigma_1,\tau_1), (p_2,\sigma_2,\tau_2)\in Q\times\Sigma^2$.
\end{enumerate}
\end{lemma}

We return to the case of $k$-tape QTMs. Now we expand the notation
$\delta[p,\sigma,\tau|\epsilon]$ as follows. Let $(p,\vsigma,\vtau)\in
Q\times (\Sigma^k)^2$ and $\vepsilon\in E^k$. Let $D_{\svepsilon}
=\{\vd\in D^k\mid \forall
i\in\{1,\ldots,k\}(|2d_i-\epsilon_i|\leq1)\}$, where $\vd=(d_i)_{1\leq
i\leq k}$ and $\vepsilon=(\epsilon_i)_{1\leq i\leq k}$. Let
$h_{\svd,\svepsilon} =(h_{d_i,\epsilon_i})_{1\leq i\leq k}$. Note that
if $|\vepsilon|\neq|\vepsilon'|$ then $h_{\svd,\svepsilon}\neq
h_{\svd',\svepsilon'}$ for any $\vd\in D_{\svepsilon}$ and any
$\vd'\in D_{\svepsilon'}$, where $|\epsilon|=(|\epsilon_i|)_{1\leq
i\leq k}$. 

We define $\delta[p,\vsigma,\vtau|\vepsilon]$ as follows:
\[
 \delta[p,\vsigma,\vtau|\vepsilon] = \sum_{q\in Q} \sum_{\svd\in
D_{\svepsilon}} \frac{\delta(p,\vsigma,q,\vtau,\vd)}{\sqrt{|E_{\svd}|}}
\qubit{q} \qubit{h_{\svd,\svepsilon}}.
\]

Since any two distinct tapes do not interfere, multi-tape QTMs must
satisfy the $k$ independent conditions of Lemma
\ref{lemma:well-formedness}. Thus, we obtain:

\begin{lemma}\label{lemma:general-formedness}
(Well-Formedness Lemma)\hs{2} Let $k\geq1$. A $k$-tape QTM
$M=(Q,\{q_0\},Q_f,\Sigma^k,\delta)$ is well-formed iff the following
three conditions hold.
\begin{enumerate}
\item (unit length) $\|\delta(p,\vsigma)\|=1$ for
all $(p,\vsigma)\in Q\times\Sigma^k$.

\item (orthogonality) $\delta(p_1,\vsigma_1)
\cdot\delta(p_2,\vsigma_2)=0$ for any distinct pairs
$(p_1,\vsigma_1), (p_2,\vsigma_2)\in Q\times\Sigma^k$.

\item (separability) $\delta[p_1,\vsigma_1,\vtau_1|\vepsilon]\cdot
\delta[p_2,\vsigma_2,\vtau_2|\vepsilon']=0$ for any distinct pair
$\vepsilon,\vepsilon'\in E^k$ and for any pair \linebreak
$(p_1,\vsigma_1,\vtau_1), (p_2,\vsigma_2,\vtau_2)\in
Q\times(\Sigma^k)^2$.
\end{enumerate}
\end{lemma}
\bs

At the end of this appendix, we give the proof of Lemma \ref{lemma:epsilon}.

\begin{proofof}{Lemma \ref{lemma:epsilon}}
Let $ACC$ and $REJ$ be the sets of all accepting configurations and
rejecting configurations, respectively. Let $ALL=ACC\cup REJ$. Assume
that $U_{M}\qubit{\phi}=\sum_{i\in ALL}\alpha_i\qubit{i}$ and
$U_{N}\qubit{\psi}=\sum_{i\in ALL}\beta_i\qubit{i}$.
\begin{eqnarray*} 
2|\eta_{M}(\qubit{\phi}) - \eta_{N}(\qubit{\psi})| &=& 
|\eta_{M}(\qubit{\phi}) - \eta_{N}(\qubit{\psi})| + 
|\overline{\eta}_{M}(\qubit{\phi}) - \overline{\eta}_{N}(\qubit{\psi})| \\ 
&\leq& 
\sum_{i\in ACC}\left||\alpha_i|^2 - |\beta_i|^2\right| + 
\sum_{i\in REJ}\left||\alpha_i|^2 - |\beta_i|^2\right| \\ 
&=& \sum_{i\in ALL} \left|(|\alpha_i|-|\beta_i|)(|\alpha_i|+|\beta_i|)\right| 
\\ 
&\leq& \sum_{i\in ALL}|\alpha_i-\beta_i||\alpha_i| + 
\sum_{i\in ALL}|\alpha_i-\beta_i||\beta_i| \\ 
&\leq& \left(\sum_{i\in ALL}|\alpha_i - \beta_i|^2 
\sum_{i\in ALL}|\alpha_i|^2\right)^{1/2} + 
\left(\sum_{i\in ALL}|\alpha_i - \beta_i|^2 
\sum_{i\in ALL}|\beta_i|^2\right)^{1/2} \\ 
&=& 2\left(\sum_{i\in ALL}|\alpha_i - \beta_i|^2\right)^{1/2} \\ 
&=& 2\|U_{M}\qubit{\phi} - U_{N}\qubit{\psi}\|_{2}, 
\end{eqnarray*} 
where $\overline{\eta}_{M}(x)$ is the rejection probability of $M$ on
input $x$. For the first and the second inequalities, we use
$||\alpha_i|-|\beta_i||\leq|\alpha_i-\beta_i|$ and the Cauchy-Schwartz
inequality.
\end{proofof}

\end{document}